\documentclass[preprint]{aastex}

\begin{document}

\title{Investigation of Primordial Black Hole Bursts using Interplanetary Network Gamma-ray Bursts}

\author{T. N. Ukwatta}
\affil{Director's Postdoctoral Fellow, Space and Remote Sensing (ISR-2), Los Alamos National Laboratory, Los Alamos, NM 87545, USA.}
\email{tilan@lanl.gov}

\author{K. Hurley}
\affil{University of California, Berkeley, Space Sciences Laboratory,
7 Gauss Way, Berkeley, CA 94720-7450, USA}

\author{J. H. MacGibbon}
\affil{Department of Physics, University of North Florida, Jacksonville, FL 32224, USA}

\author{the following authors are in order of the number of times their experiments were used
in the paper}

\author{D. S. Svinkin, R. L. Aptekar, S. V. Golenetskii, D. D. Frederiks, V. D. Pal'shin}
\affil{Ioffe Physical Technical Institute, St. Petersburg, 194021, Russian Federation}

\author{J. Goldsten}
\affil{Applied Physics Laboratory, Johns Hopkins University, Laurel, MD 20723, U.S.A.}

\author{W. Boynton}
\affil{University of Arizona, Department of Planetary Sciences, Tucson, Arizona 85721, U.S.A.}

\author{A. S. Kozyrev}
\affil{Space Research Institute, 84/32, Profsoyuznaya, Moscow 117997, Russian Federation}

\author{A. Rau, A. von Kienlin, X. Zhang}
\affil{Max-Planck-Institut f\"{u}r extraterrestrische Physik,
Giessenbachstrasse, Postfach 1312, Garching, 85748 Germany}

\author{V. Connaughton}
\affil{ University of Alabama in Huntsville, NSSTC, 320 Sparkman Drive, Huntsville, AL 35805, USA}


\author{K. Yamaoka}
\affil{Department of Physics and Mathematics, Aoyama Gakuin University, 5-10-1 Fuchinobe, Sagamihara, Kanagawa 229-8558, Japan}

\author{M. Ohno}
\affil{Department of Physics, Hiroshima University, 1-3-1 Kagamiyama, Higashi-Hiroshima, Hiroshima 739-8526, Japan}

\author{N. Ohmori}
\affil{Department of Applied Physics, University of Miyazaki, 1-1 Gakuen kibanadai-nishi, Miyazaki-shi, Miyazaki 889-2192, Japan}





\author{M. Feroci}
\affil{INAF/IAPS-Roma, via Fosso del Cavaliere 100, 00133, Roma, Italy}

\author{F. Frontera\altaffilmark{1}, C. Guidorzi}
\affil{University of Ferrara, Dept. of Physics and Earth Science, via Saragat 1, 44122 Ferrara, Italy}
\altaffiltext{1}{INAF/Istituto di Astrofisica Spaziale e Fisica Cosmica di Bologna, via Gobetti 101, I-40129 Bologna, Italy}

\author{T. Cline\altaffilmark{2}, N. Gehrels}
\affil{NASA Goddard Space Flight Center, Code 661, Greenbelt, MD 20771, U.S.A.}
\altaffiltext{2}{Emeritus}
\altaffiltext{3}{Joint Center for Astrophysics, University of Maryland, Baltimore County, 1000 Hilltop Circle, Baltimore, MD 21250}
\altaffiltext{4}{Universities Space Research Association, 10211 Wincopin Circle, Suite 500, Columbia, MD 21044}


\author{H. A. Krimm\altaffilmark{4}}
\affil{USRA/CRESST/NASA Goddard Space Flight Center, Code 661, Greenbelt, MD 20771, U.S.A.}


\author{J. McTiernan}
\affil{University of California, Berkeley, Space Sciences Laboratory, 7 Gauss Way, Berkeley, CA 94720-7450, USA}

\begin{abstract}
The detection of a gamma-ray burst (GRB) in the solar neighborhood would have
very important implications for GRB phenomenology. The leading theories for
cosmological GRBs would not be able to explain such events. The final bursts of
evaporating Primordial Black Holes (PBHs), however, would be a natural explanation
for local GRBs. 
We present a novel technique that can constrain the distance to gamma-ray bursts using detections from widely separated, non-imaging spacecraft. This method can determine the actual distance to the burst if it is local.
We applied this method to constrain distances to a sample of 36 short duration GRBs 
detected by the Interplanetary Network (IPN) that show
observational properties that are expected from PBH evaporations.
These bursts have minimum possible distances in the 
10$^{13}$--10$^{18}$ cm (7--10$^5$ AU) range, consistent with
the expected PBH energetics and with a possible origin in the solar neighborhood, 
although none of the bursts can be unambiguously demonstrated to be local. 
Assuming these bursts are real PBH events,
we estimate lower limits on the PBH burst evaporation
rate in the solar neighborhood.
\end{abstract}


\section{Introduction}\label{introduction}

The composition of the short-duration, hard-spectrum gamma-ray burst (GRB) population
is not yet fully understood. It is believed that most of the bursts are generated in
compact binary mergers \citep{e1} and while the handful of optical counterparts and
host galaxies discovered to date does not contradict this view, it is also thought
that the population probably contains up to 8\% extragalactic giant magnetar flares
as well \citep{h2,m1,s1}.  For the majority of the short-duration GRB population,
however, there is simply not enough evidence to determine their origin unambiguously.
Hawking radiation from primordial black holes (hereafter PBH) was one of the very first
explanations proposed for cosmic gamma-ray bursts \citep{h1},
and it continues to be proposed today \citep{c1,c2,c3,c4,c5,c6,c7}. 
The PBH lifetime and burst duration depend on its mass, so PBHs bursting today 
have similar masses and durations, and release similar energies, making them in essence
`standard candles'. The typical PBH gamma-ray burst is not expected to be 
accompanied by detectable intrinsically-generated 
extended emission or have an afterglow, although accompanying
bursts at other wavelengths or afterglows may arise if, for example, the PBH is embedded
in a high density magnetic field or plasma \citep{mcp,r1,j1}.
In the standard emission scenario, so-called because it uses the Standard Model of particle physics \citep{mw}, the PBH gamma-ray burst is strongest in the
final second of the burst lifetime, has a hard energy spectrum, and should be detectable
in the vicinity of the Earth. 
For a typical interplanetary network detector
sensitive to bursts of fluence 10$^{-6}$ erg cm$^{-2}$ and above, PBH events could
in principle be detected out to a distance of a few parsecs, depending on the emission
model. PBHs evaporating today do not have enough luminosity to be detected
at cosmological distances even by the most sensitive current instruments, so searching
for them locally is a logical step.

When observed by a single detector, the properties of a PBH burst might not appear to be
significantly different from those of other short bursts; instruments with
localization or imaging capabilities would obtain their arrival directions as they
would for an infinitely distant source.  Indeed many attempts to find evidence for
the existence of PBH bursts have to date been based mainly on the spatial
distribution and time histories of a subset of short bursts \citep{c1,c2,c3,c4,c5,c6,c7}.
Other search methods have employed atmospheric Cherenkov detectors 
\citep{pw1,pw2,pw3,linton2006,veritas2012,glicenstein2013},
air shower detectors \citep{fe1,bh1,alex1993,amenomori1995,abdo2015}, 
radio pulse detection \citep{pt1,ke1}, spark
chamber detection \citep{fi1}, and GRB femtolensing \citep{ba1}. Table~\ref{comparison}
gives a comparison of these various methods.

To widely spaced interplanetary network (IPN) detectors, however, a local PBH burst
could look significantly different when compared with bursts from distant sources, due
to the curvature of the received wavefront. In this paper, we use this fact to explore
the possibility that some short bursts may originate in the solar neighborhood, and
estimate lower limits to the PBH burst evaporation rate assuming these bursts 
are real PBH bursts. This paper is organized as follows. In Section \ref{blackhole} we
derive the fluence expected in the detector from a PBH burst using the standard emission
model and, as a maximal alternative, the Hagedorn emission model.
In Section \ref{methodology} we explain
how we localized the detected bursts in 3D relaxing the assumption that they are at
infinite distances. The detailed discussion of the methodology is given in Appendix A.
Our data selection criteria are described in Section~\ref{data_selection}. Our results and
PBH burst rate limit calculation are given in Section~\ref{results}. 
In Section~\ref{discussion}, we discuss implications and limitations of our results.

\section{PBH Burst Signatures}\label{blackhole}

As a PBH Hawking-radiates, its mass is decreased by the total energy carried off by the
emission, and the black hole temperature, which is inversely
proportional to its mass, increases. In the standard emission model (SEM)~\citep{mw}, the
black hole directly Hawking-radiates those particles which appear fundamental on the
scale of the black hole. Once the black hole temperature reaches the QCD
transition scale ($\sim 200 - 300$ MeV), quarks and gluons are directly Hawking-radiated.
The PBH gamma-ray burst spectrum is the combination of the directly Hawking-radiated
photons and those produced by the decay of other directly Hawking-radiated particles. 
An SEM PBH with a remaining emission lifetime of
$\tau\lesssim 1$ sec has a mass of
\begin{equation}\label{eq:masstau}
M_{BH}(\tau) \approx 1.3 \times 10^{9} \left(\frac{\tau}{1 \ \rm s}\right)^{1/3}\ {\rm g}
\end{equation}
\citep{ukwatta2015pbh} and a remaining rest mass energy of
\begin{equation}\label{eq:energytau}
E_{BH}(\tau) \approx 1.2 \times 10^{30} \left(\frac{\tau}{1 \ \rm s}\right)^{1/3}\ {\rm erg}.
\end{equation}
The expected fluence arriving at the detector from a PBH at a distance $d$ from Earth 
is then
\begin{equation}\label{eq:semfluence}
F_{D}=\frac{E_{\gamma BH}}{4\pi d^2}
\end{equation}
where $E_{\gamma BH} = \eta_{\gamma D} E_{BH}$ and $\eta_{\gamma D}$ is the fraction of
the PBH energy that arrives in the energy band of the detector, and the maximum distance
from which the SEM PBH is detectable is
\begin{equation}\label{eq:semdmax}
d_{max}\simeq 0.01\left(\frac{\eta_{\gamma D}}{10^{-2}}\right)^{1/2}\left(\frac{\tau}{1\ \rm{s}}\right)^{1/6}\left(\frac{F_{D\ min}}{10^{-6}\ \rm{erg}\ \rm{cm}^{-2}}\right)^{-1/2}\ \rm{pc}
\end{equation}
where $F_{D\ min}$ is the sensitivity of the detector.

The SEM analysis is consistent with high energy accelerator experiments~\citep{mcp}.
However, an alternative class of PBH evaporation models was proposed before the existence
of quarks and gluons was confirmed in accelerator experiments and these models continue
to be discussed in the PBH burst literature. In such models (which we label HM
scenarios), a Hagedorn-type exponentially increasing number of degrees of freedom become
available as radiation modes once the black hole temperature reaches a specific threshold
such as the QCD transition scale. In the HM scenarios, we assume that the remaining PBH
mass is emitted quasi-instantaneously as a burst of energy $E'_{BH} = M'_{BH}c^2$ once
the black hole mass reaches some threshold $M'_{BH}$; for the QCD transition scale,
$M'_{BH}\sim 10^{14}$ g and $E'_{BH}\sim 10^{35}$ erg. Proceeding as above, the maximum
distance from which the HM PBH burst is detectable is
\begin{equation}\label{eq:hdmax}
d'_{max}\simeq 9\left(\frac{\eta'_{\gamma D}}{10^{-1}}\right)^{1/2}\left(\frac{M'_{BH}}{10^{14}\ \rm{g}}\right)^{1/2}\left(\frac{F_{D\ min}}{10^{-6}\ \rm{erg}\ \rm{cm}^{-2}}\right)^{-1/2}\ \rm{pc}
\end{equation}
where $\eta'_{\gamma D}$ is the fraction of the HM PBH energy that arrives in the energy band of the detector.

\section{Gamma-ray burst localization for a source at a finite distance}\label{methodology}

When a pair of IPN spacecraft detects a burst, if the distance to the source is taken
to be a free parameter, the event is localized to one sheet of a hyperboloid of revolution about
the axis defined by the line between the spacecraft.  If the burst is assumed to be
at a distance which is much greater than the interspacecraft distance, the hyperboloid
intersects the celestial sphere to form the usual localization circle (or annulus,
when uncertainties are taken into account). Another widely spaced spacecraft
would produces a second hyperboloid which intersects the first one to define a locus of
points which is a simple hyperbola. Note that both hyperboloids have a common focus.
Again, if the burst is assumed to be at a large
distance from the spacecraft, the hyperbola intersects the celestial sphere at two
points to define two possible error boxes. A fourth, non-coplanar spacecraft even
at a moderate distance from Earth, such as \it Konus-WIND, \rm can often be used to
eliminate one branch of the hyperbola and part of the second branch. A terrestrial analogue to this method is Time Difference of Arrival (TDOA), with the important exception that GRB sources can be at distances which are effectively infinite. Further
details are given in Appendix A.  While a single instrument with imaging or
localization capability would obtain the correct sky position for a PBH burst regardless
of its distance, the same is not true of an IPN localization, for which the derived
arrival direction depends on the source distance.

In this paper, we relax the assumption that bursts are at
infinite distances. If a burst is detected by three widely
spaced spacecraft, then according to the previous discussion, the possible location
of the burst traces a simple hyperbola in space as illustrated in Figure~\ref{fig:figure3}. In an Earth-centered
coordinate system, this hyperbola has a closest distance to the Earth, that is, a distance
lower limit. As we
explain in Section~\ref{ul_estimate}, this fact can be used to calculate a
lower limit to the PBH burst density rate in the Solar neighborhood assuming
that the bursts that we consider are actual PBH bursts. In principle,
detections by three spacecraft can rule out a local origin for a burst, but it is impossible to prove a local origin with only three non-imaging spacecraft.

In the case where a burst is observed by four widely
spaced non-imaging spacecraft, the burst can be localized to a single point in space (or a region in space if uncertainties are taken into consideration). This scenario is illustrated in Figure~\ref{fig:figure4}. Thus in order to prove the local origin of a burst using non-imaging spacecraft, one needs detections from at least four satellites 
that are at interplanetary distances from each other.

As mentioned in Appendix A, in the special case of two widely separated spacecraft, where one spacecraft has precise imaging capability, it is in principle possible to demonstrate a local origin.  We have explored this case in detail and defer treatment of it to a future paper.  None of the events in this paper are in that category.

\section{Data Selection}\label{data_selection}

The IPN database contains information on over 25,000 cosmic, solar, and magnetar
events which occurred between 1990 and the present (\texttt{http://ssl.berkeley.edu/ipn3/index.html}). During this period, a total of 18 spacecraft participated in the network.  Some were dedicated GRB detectors, while others were primarily gamma-ray detectors with GRB detection capability.  Indeed, the composition of the IPN changed regularly during this time, as old missions were retired and new ones were launched.  However, all the instruments were sensitive to bursts with fluences around $\rm 10^{-6} ~erg ~cm^{-2}$ or peak fluxes $\rm ~1 ~photon ~cm^{-2} ~s^{-1}$ and above, resulting in a roughly constant detection rate.
All known bursts, regardless of their intensity or duration, or the instruments
which detected them, are included in this list.  We have searched it for gamma-ray
bursts with the following properties.
\begin{itemize}
\item Confirmed cosmic bursts which occurred between 1990 October and
2014 December (24.25 y; 10795 GRBs survived this cut).
\item Bursts observed by three or more spacecraft, of which two were at interplanetary
distances; 839 GRBs survived this cut. This small number is due firstly to the relatively high sensitivity thresholds of the distant IPN detectors
(roughly $\rm 10^{-6} ~erg ~cm^{-2} ~or ~1 ~photon ~cm^{-2} ~s^{-1}$),
and their somewhat coarser time resolutions, and secondly to a 2.5 year period between 1993 August and 1996 February when there was only one interplanetary spacecraft in the network. 
\item Bursts with no X-ray or optical afterglow, either because there were no
follow-up observations, or because searches were negative. 
In addition, as discussed before, the arrival
direction derived from IPN localization depends on its assumed distance, and the
bursts were initially triangulated assuming an infinite distance.  Thus even
if a simultaneous search had taken place, it might not
have identified an event within the error box if the burst was local. Other selection effects come into play starting with the launch of the HETE spacecraft in 2000 October, and later with the launch of Swift in 2004 November, namely that X-ray and optical observations were often done rapidly, 
leading to more X-ray and optical detections and the elimination of the bursts from further consideration here. On the other hand, the launches of Suzaku in 2005 July and Fermi in 2008 June resulted in an increase in the short burst detection rate which more than compensated for the previous effect.  
\item Bursts with durations $<$ 1 s and no extended emission (EE). No cut was made
based on the light curve shape; we discuss this in section~\ref{discussion}. The detection rate of short bursts was roughly, but not exactly, constant over the period of this study.
\end{itemize}

These cuts, which are commutative, were carried out in the order described
above, so as to minimize the required analysis of the full sample of 10795 events.
36 bursts satisfied these criteria. None were detected by an imaging instrument with good spatial resolution. While we believe that the overall IPN detection rate of short bursts was roughly constant, the data selection resulted in a rate which varied significantly from year to year.

In order to obtain distance lower limits as described in Appendix A, these bursts
were triangulated assuming that their distances from Earth were free parameters.
In all cases, however, an infinite distance was
also compatible with the data.  The results appear in Table~\ref{distances}.

Of the 10795 GRBs which survived the first cut, we would expect roughly 20\%,
or 2150, to be short-duration events ($<$ 1 s).  Of those, perhaps 10\%, or 215,
would display extended emission \citep{pa2}, bringing the sample to 1935 non-EE
bursts.  We would expect about 80\% of them to have no optical counterpart,
either because none was detected or none was searched for (this number applies
only to short bursts).  This reduces the sample to
about 1550.  Since only
36 events survived all the data cuts, we can estimate the average IPN efficiency for
this selection procedure to be about 2.3\%.  Thus if we exclude the 2.5 year period when the IPN had a single interplanetary spacecraft, the observed event rate  is
1.7/year, and the true rate is about 72/year. 

\section{Results}\label{results}

\subsection{Distance Limits and Localizations of PBH Burst Candidates}

According the methodology described in Section~\ref{methodology} and
Appendix A, we have calculated the minimum possible distances to the
sample of 36 bursts selected in section~\ref{data_selection}. This
burst sample is shown in Table~\ref{distances} and the 12 columns give:

\begin{enumerate}
\item the date of the burst, in \texttt{yymmdd} format, with suffix
A or B where appropriate,
\item the Universal Time of the burst at Earth, in seconds of day,
\item the spacecraft which were used for the triangulation; a complete list
of the spacecraft which detected the burst may be
found on the IPN website (\texttt{http://ssl.berkeley.edu/ipn3/index.html}),
\item the burst duration, in seconds,\rm
\item the fluence of the burst in erg cm$^{-2}$,
\item the energy range over which the duration and fluence were measured, in keV,
\item the lower limit to the burst distance, obtained by triangulation, in cm,
\item the distance to which this burst could have been detected if it were a PBH
burst of energy 10$^{34}$ erg, assuming that all the energy went into the
measured fluence (this is essentially the maximum possible detectable distance),
\item the maximum detectable distance assuming the SEM model (Equation~\ref{eq:semdmax})
in terms of the undetermined parameter $(\eta_{\gamma D})^{0.5}$,
\item the maximum detectable distance assuming the HM model (Equation~\ref{eq:hdmax})
in terms of the undetermined parameter $(\eta'_{\gamma D})^{0.5}$,
\item whether or not counterpart searches took place and if so, their references,
\item references to the duration, peak flux, and/or fluence measurements, and/or
to the localization.
\end{enumerate}

The shortest burst in Table~\ref{distances} has a duration of 60 ms.  Due to
the relatively coarse time resolutions of interplanetary detectors, bursts
with shorter durations must have greater intensities to be detected, effectively setting
a higher detection threshold for very short events.  The weakest event has
a fluence of $4.65\times 10^{-7} \, \rm erg \, cm^{-2}$. The bursts in
Table~\ref{distances} could not have come from distances less than the distance
lower limits in column 7; however, all of them have time delays which are also consistent
with infinite distances. Figure~\ref{fig:pbh_min_dist_histo} shows a histogram of 
these minimum distances. 
The detector-dependent distance upper limits in column 8 are calculated assuming 
the extreme case that these events are caused by $\sim 10^{34}$ erg HM-type bursts 
from primordial black holes of mass $\sim 10^{14}$ g and that all of the emitted 
energy spectrum is contained within the detector measurement limits.
Table~\ref{localizations} gives the coordinates of the centers and
corners of the error boxes for the events in  Table~\ref{distances}, assuming that the
sources are at infinity.  If in fact the sources are local, the arrival directions
are distance-dependent, and different from the ones in Table~\ref{localizations}.
These coordinates represent the intersections of annuli, and in some cases the curvature
of the annuli would make it inaccurate to construct an error box by connecting the
coordinates with straight-line segments.

\subsection{PBH Burst Density Rate Estimation}\label{ul_estimate}

All previous direct PBH burst searches resulted in null detections \citep{abdo2015}. In this case, one can derive an upper limit on the local PBH burst rate density, that is, an upper limit on the number of PBH bursts per unit volume per unit time in the local solar neighborhood.

However, in our case, we have PBH burst candidates with short duration, no known afterglow detection and 
minimum distances that are sub-light-years. Since we have PBH candidates, we should be able to derive an actual 
measurement of the PBH burst rate density under the assumption that the candidates are actual PBH bursts. Thus,
the actual PBH burst rate density is
\begin{equation}\label{pbh_rate}
R = \frac{n}{V S \epsilon}
\end{equation}
where $n$ is the number of PBH bursts, $V$ is the effective PBH detectable volume and $S$ is the observed 
duration. The selection efficiency of the IPN is $\epsilon$.

If all the candidates identified in Section~\ref{data_selection} are real PBH bursts, then we have 36 
PBH bursts, i.e., $n=$36. Hence, our PBH burst rate density estimate is,
\begin{equation}\label{pbh_rate_36}
R = \frac{36}{V S \epsilon}.
\end{equation}

Next we need to estimate the values of $S$, $V$, and $\epsilon$. Because we have 
studied IPN bursts collected over 21.75 years (the 2.5 year IPN non--sensitivity period is excluded), our observed duration is $S$=21.75 years. The effective PBH detectable volume, $V$, calculation for this 21.75 year period, however, is
not obvious. Each PBH candidate has a distance consistent with some minimum distance up to
infinity. We also know that PBH bursts are not bright enough to be detected from large
distances. The maximum possible detectable distance of a PBH burst depends on the
high-energy physics model used to calculate the final PBH burst
spectrum~\citep{ukwatta2015pbh}. Currently there are no accurate calculations for final PBH
burst photon spectra in the keV-MeV energy range. Thus as a conservative maximum possible
detectable PBH burst distance, we can take the maximum value of the minimum distances in
our candidate PBH burst sample. This corresponds to a distance of 0.47 parsecs ($1.5 \times
10^{18}$ cm). 
Because all the PBH burst candidates in the sample are actual IPN detections, this distance
value is model-independent. On the other hand, it is important to note that the IPN is not
capable of detecting all the bursts
within this distance over the entire observation duration due to various factors such as
the orientation of the satellites, and/or instrument duty cycles. Thus the effective volume
calculated from the above maximum possible detectable PBH burst distance is an
overestimate. Hence the PBH burst rate calculated in Equation~\ref{pbh_rate_36} is in reality a
lower limit on the PBH burst rate density,
\begin{equation}\label{pbh_rate_ll}
R_{LL} = \frac{36}{V S \epsilon}.
\end{equation}

In Section~\ref{data_selection}, we made a rough estimate of the selection efficiency of
IPN, $\epsilon$, for PBH bursts. However, we note that it is very challenging to calculate $\epsilon$ accurately due to a number of unknown factors such as the fraction of bursts without EE, the fraction of bursts without afterglows, the fraction of bursts to which the IPN is not sensitive (for example due to orientation or deadtime), etc.

Using the estimated effective PBH detectable volume $V$, observed duration $S$, and selection efficiency $\epsilon$, 
we can now estimate the lower limit of the PBH burst rate in the best case scenario where all the PBH
burst candidates are actual PBH bursts. In this case, our PBH burst lower limit is $\sim$158.5 bursts
$\rm pc^{-3}\,yr^{-1}$. If we assumed 100\% efficiency then the PBH burst lower limit is $\sim$ 3.6 bursts $\rm pc^{-3}\,yr^{-1}$.
If only one of the candidates is an actual PBH burst then the value of the rate density
lower limit depends on the minimum distance to that particular burst. If the burst with the
largest minimum distance (GRB 140807) is the PBH burst, then the PBH burst rate
density lower limit is $\sim$ 0.1 bursts
$\rm pc^{-3}\,yr^{-1}$. If the burst with the smallest minimum distance (GRB 970902)
is real then the PBH burst rate density lower limit is $\sim \ 1 \times 10^{14}$ bursts
$\rm pc^{-3}\,yr^{-1}$ (this value is excluded by other high-energy experiments, however).
All these estimates assume that PBHs are distributed uniformly in the solar neighborhood.
The IPN PBH burst rate density lower limit values are shown in Figure~\ref{fig:pbh-limit-final}.
PBH burst upper limits from various other searches are also shown in the figure.

In the worst case scenario where none of our candidates is a real PBH burst, we cannot
estimate a lower limit to the PBH burst rate density and instead consider 
to estimate an upper limit to the PBH burst rate density. However, the assumption that none of the bursts in the sample is real but still we have candidates implies that our criteria to identify PBH bursts defined in section~\ref{data_selection} is not sufficient. This means our method is not capable of setting an upper limit on the PBH burst rate density.

\section{Discussion}\label{discussion}

The detection of gamma-ray transients points to very high energy explosive phenomena in the
Universe. Their detection in the solar neighborhood would indicate a previously 
unrecognized and potentially exotic phenomenon in our cosmic backyard. The sample of 
bursts identified in this paper are candidates for such explosions. They have short 
durations, no known afterglow detections, and have distance limits consistent with the solar neighborhood. In principle our methodology is capable of proving 
the local origin of bursts. However in order to do that we need either
four widely separated non-imaging spacecraft or two spacecraft that include one with imaging capability (see Section~\ref{methodology} and Appendix A). This is not the case for any of the bursts we considered in our study. While some events were indeed detected by four or more spacecraft, the spacecraft were not widely separated, i.e. at interplanetary distances. With four widely separated non-imaging spacecraft detections or detections by one imaging spacecraft and one non-imaging spacecraft, it would be possible to prove that some of these bursts 
are in the solar neighborhood and this would definitely point to an exotic  origin for these bursts. Lacking that however, we can look at other properties of 
these bursts and discuss how likely it is that they may have a PBH origin.

Firstly, it is of interest to investigate the sky distribution
of our PBH burst candidates. For example,~\cite{c3} have argued that, due to the fact 
that very short duration GRBs (i.e., GRBs with duration
$\le$ 100 msec) have a non-isotropic sky distribution, they may be drawn from a
different GRB population, possibly from PBH bursts.  
In order to investigate this we have calculated burst density maps 
using the Gaussian kernel density methodology described in \cite{ukwatta2015}.
Since the sky locations of the PBH candidates depends on their distance from 
earth, we started by assuming all the bursts are at their minimum distances and calculated 
their sky density map. This map is shown in Figure~\ref{sky_map_pbh_candidate_min}. 
The map is presented in Galactic coordinates with a 25 degree smoothing radius. It shows some relatively high density areas, but
the probability of generating this density contrast by chance, in the case when the 
true sky distribution is uniform, is $\sim$0.2, estimated using a Monte Carlo 
simulation  ~\citep{ukwatta2015}. Thus the density structure
seen in Figure~\ref{sky_map_pbh_candidate_min} is consistent with a uniform source 
distribution. If these PBH candidates are real, they cannot be further away than $\sim$10 parsecs
(which is the maximum possible distance they can be detected assuming the optimistic Hagedorn-type model). We then also calculated the sky locations of the PBH candidates assuming they are at
10 parsecs and derived the sky density map as shown in Figure~\ref{sky_map_pbh_candidate_10pc}. This map is also consistent with a uniform source 
distribution. This is the behaviour one would expect if these PBH burst are local with
maximum detectable distance $\sim$10 parsecs.

According to the standard model for Hawking radiation \citep{mw}, PBH bursts are 
standard candles, that is, all PBH bursts are intrinsically identical at the source. 
However, the way the burst appears at large distances
may vary depending on its host environment. The final burst properties
of the PBH burst depend on its mass and the number of fundamental particle degrees 
of freedom available at various
energies~\citep{ukwatta2015pbh}. In principle, by measuring the photon flux 
arriving from a PBH burst candidate in a given energy range, we can calculate the 
distance to that burst. However, during the last second of the PBH lifetime, its 
temperature is well above $\sim$ 1 TeV and the physics governing these high energies 
is not fully understood. Thus, the full spectrum of a PBH burst is difficult to 
calculate.

\cite{ukwatta2015pbh} have calculated PBH burst light curves in the 50 GeV -- 100 TeV 
energy range using the Standard Model of particle physics. This calculation considered 
both the direct Hawking radiation of photons from the black hole and the photons created
due to the fragmentation and hadronization of the directly Hawking-radiated quarks and
gluons. The expected PBH light curve is a power-law
with an index of $\sim 0.5$ and at various sub-energy bands
within the 50 GeV -- 100 TeV energy range has an interesting inflection
point due to the brief dominance of the directly Hawking-radiated photons 
around $\sim 0.1$ second before the PBH expiration. This is potentially a
unique signature of a PBH burst in the GeV/TeV energy range.

The detectors in the IPN are sensitive to photons in the energy range 
10 keV -- 100 MeV. There are no published calculations of PBH burst light curves 
in the keV/MeV range using the Standard Model of particle physics to which we can fit our light curves and extract fit parameters. Nonetheless it is of
interest to look at the light curves of our PBH burst candidates. 
Firstly, we note that, although a PBH will be emitting in the keV/MeV range before it becomes hot enough to emit in the GeV/TeV range, a keV/MeV burst signal can only be generated by the low energy component of a PBH that is also emitting in the GeV/TeV range: this is because a 10 TeV black hole has a remaining burst lifetime of $\sim$1 s whereas a $<$1 GeV black hole has a lifetime $\gg 10^4$ yr and observationally would be a stable source not a burst. This low energy photon component, in turn, is predominantly generated by other higher energy Hawking-radiated species via decays or the inner bremsstrahlung effect~\citep{Page2008} and is not the directly Hawking-radiated photon flux which decreases in the keV/MeV band as the burst progresses. Acknowledging the uncertainty in the PBH light curves in the keV/MeV range, it is also possible that the PBH burst 
signal may have a longer or shorter duration in the keV/MeV range than in
the GeV/TeV range due to differences both in production at the source and in detector sensitivity, and that the duration difference varies with the distance to the PBH. Figure~\ref{lc_grid} shows the light curves of the 36 IPN bursts in our sample. Some are clearly single-peaked, others are clearly multi-peaked, and some were not recorded with sufficient statistics to determine the true number of peaks. It is interesting to note that bursts such as GRB 970921, GRB 080222, GRB000607, GRB 101009, GRB 121127, GRB 131126A, and GRB 141011A display a keV/MeV time profile that resembles the PBH light curve profile calculated by~\cite{ukwatta2015pbh} for the GeV/TeV energy range.

In this paper, we introduced a novel method to constrain the distances to GRBs using
detections from multiple spacecraft. Utilizing detections from three non-imaging spacecraft we could only constrain the minimum distances to our current sample of bursts.
The maximum distance is constrained 
by the energy available during the final second of the PBH burst. However, the
amount of energy released in the keV/MeV energy band is not known and may be 
highly model-dependent. On the other hand, with detections by four widely separated ($\sim$ AU distances) non-imaging
spacecraft or one non-imaging spacecraft and one imaging spacecraft, 
we can constrain burst distances
independent of any high energy physics model, and potentially show that
some bursts are local. Such a detection will not only
prove the existence of PBH bursts, by fitting light curves and spectra 
derived using various beyond the standard model physics theories, we 
can also identify which theory describes nature.

\section{Acknowledgments}

Support for the IPN was provided by NASA grants NNX09AU03G, NNX10AU34G, NNX11AP96G, and NNX13AP09G (\it Fermi\rm); NNG04GM50G, NNG06GE69G, NNX07AQ22G, NNX08AC90G, NNX08AX95G, and NNX09AR28G
(INTEGRAL); NNX08AN23G, NNX09AO97G, and NNX12AD68G(\it Swift\rm); NNX06AI36G, NNX08AB84G, NNX08AZ85G, NNX09AV61G, and NNX10AR12G (\it Suzaku\rm); NNX07AR71G (MESSENGER);
NAG5-3500, and JPL Contracts 1282043 and Y503559 (\it Odyssey\rm); NNX12AE41G, NNX13AI54G, and NNX15AE60G (ADA);
NNX07AH52G (Konus); NAG5-13080 (RHESSI);
NAG5-7766, NAG5-9126, and NAG5-10710,  (\it BeppoSAX \rm); and
NNG06GI89G.
TNU acknowledges support from the Laboratory Directed Research and Development
program at the Los Alamos National Laboratory (LANL).
The Konus-Wind experiment is partially supported by a Russian Space Agency contract and RFBR grants 15-02-00532 and 13-02-12017-ofi-m.
We would also like to thank Jim Linnemann (MSU), Dan Stump (MSU),
Brenda Dingus (LANL), and Pat Harding (LANL) for useful conversations on
the analysis.

\appendix
\section{GRB triangulation when the source distance is allowed to vary}

Assume two spacecraft, $SC_1$ and $SC_2$, separated by a distance $d$, observe a GRB.
For any assumed distance between the GRB and the spacecraft,
the difference in arrival times $\delta$t$_{12}$ must be constant.  Let
$x_1$, $y_1$, $z_1$ and $x_2$, $y_2$, $z_2$ be the coordinates of the two
spacecraft.  Then the locus of points $x,y,z$ which describes the possible
source locations is given by

\begin{eqnarray*}
\sqrt{(x-x_1)^2 + (y-y_1)^2 + (z-z_1)^2} - \sqrt{(x-x_2)^2 + (y-y_2)^2 + (z-z_2)^2} - c*\delta t_{12} =0
\end{eqnarray*}

where $c$ is the speed of light.

Consider first the two-dimensional problem for simplicity.
Let $SC_1$ and $SC_2$ define the z-axis of a coordinate system whose origin is halfway between the
spacecraft. The positions of $SC_1$ and $SC_2$ are the foci of the hyperbola:

\begin{eqnarray*}
z^2/a^2 - x^2/b^2=1
\end{eqnarray*}

\begin{figure}
\begin{center}
\includegraphics[width=0.8\textwidth]{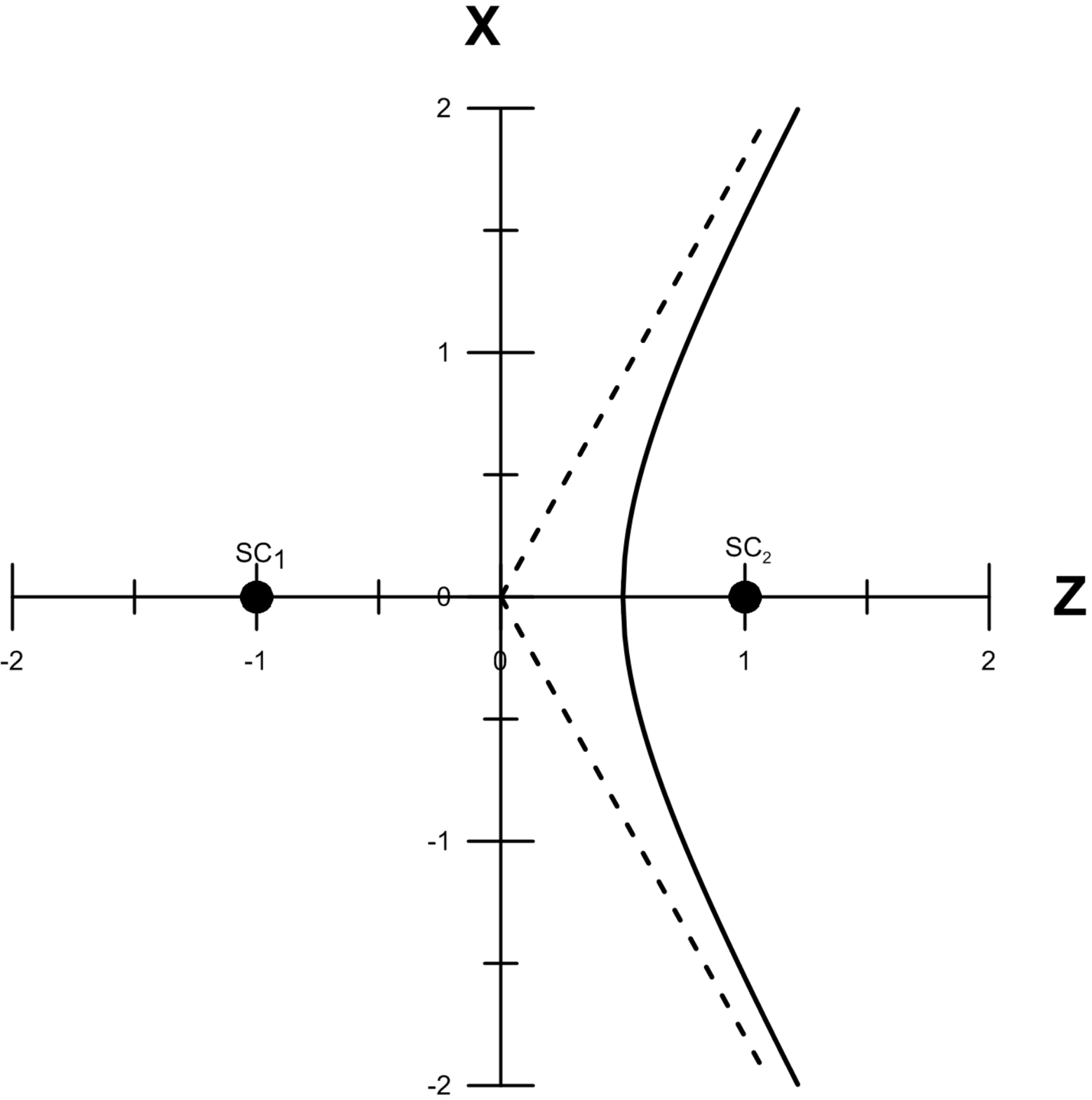}
\caption{A two-dimensional example of GRB triangulation when the source distance is allowed
to vary.  The two spacecraft, 1 and 2, are aligned along the z-axis, and are the foci of a hyperbola.  The hyperbola defines the loci of
possible source distances.  If the distance is assumed to be infinite, the two possible GRB arrival directions are along the asymptotes (dashed lines).
\label{fig:hyperbola}}
\end{center}
\end{figure}

This is shown in Figure~\ref{fig:hyperbola}. Here $2a$ is the difference between the distances of any point on the hyperbola from the foci, 
so $2a=c \, \delta t_{12}$, and $b=(d^2/4-a^2)^{1/2}$.
For every point on this hyperbola, the difference in the arrival times is $\delta t_{12}$.  If we assume an infinite distance for the source, the asymptotes of the hyperbola
define the two possible arrival directions of the GRB.

Now consider the three-dimensional case.  If we rotate the hyperbola of Figure 1 about the z axis, we obtain one sheet of a hyperboloid of rotation of two sheets.  Its formula is

\begin{eqnarray*}
-x^2/b^2 + z^2/a^2 - y^2/b^2=1.
\end{eqnarray*}

\begin{figure}
\begin{center}
\includegraphics[width=0.8\textwidth]{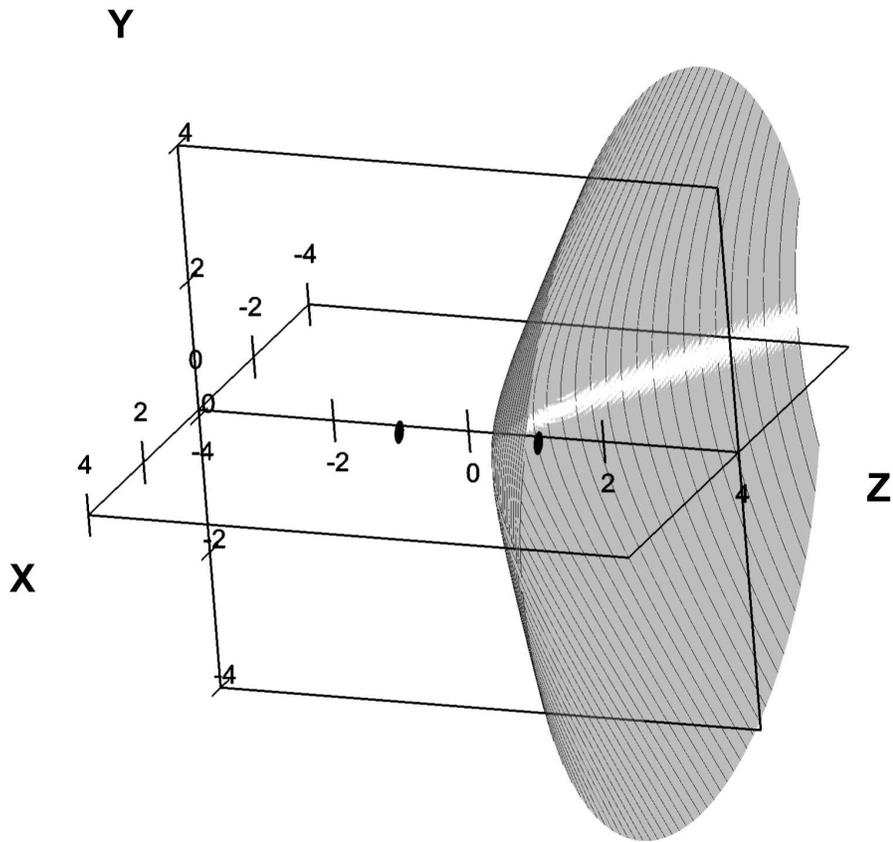}
\caption{The hyperbola of Figure~\ref{fig:hyperbola} rotated to obtain one sheet of a hyperboloid of rotation of two sheets. The two spacecraft are aligned along the z-axis, and are the foci of the hyperboloids. Each hyperboloid defines the loci of
possible source distances in the three-dimensional problem.  If the distance is assumed to be infinite,
the circle defined by the intersection of the hyperboloid with the celestial sphere gives the possible GRB arrival directions.
\label{fig:hyperboloid}}
\end{center}
\end{figure}

Here, the x axis is perpendicular to the y and z axes, and cuts in the plane z=constant give circles.  This is illustrated in Figure~\ref{fig:hyperboloid}.

In practice, we will have two or more hyperboloids generated by three or more spacecraft, and we will want to work in Earth-centered Cartesian coordinates with one axis
oriented towards right ascension zero, declination zero, and another axis oriented towards declination 90$\arcdeg$.  Consider the three-spacecraft case.  A spacecraft pair will define two foci of a hyperboloid; the line
joining the two spacecraft, which defines the axis of rotation of the hyperboloid, will be oriented with respect to Earth-centered coordinates such that it represents
a rotation and a translation.  We want to express the formula for the hyperboloid in the Earth-centered system.

The coordinate rotation can be described by three sets of direction cosines:

\begin{eqnarray*}
z^\prime=x*\cos(\alpha z_{12}) + y*\cos(\beta z_{12}) + z*\cos(\gamma z_{12})\\
y^\prime=x*\cos(\alpha y_{12}) + y*\cos(\beta y_{12}) + z*\cos(\gamma y_{12})\\
x^\prime=x*\cos(\alpha x_{12}) + y*\cos(\beta x_{12}) + z*\cos(\gamma x_{12})
\end{eqnarray*}

Here the primed coordinate system is the one defined by the foci of the hyperboloid; its origin is the same as that of the
unprimed, Earth-centered system, and it is rotated, but not translated, with respect to it.  Now perform a translation of
the primed system so that its origin is at the midpoint of the two foci.  If the coordinates of the two spacecraft, expressed
in the unprimed system, are $x_1$, $y_1$, $z_1$ and $x_2$, $y_2$, $z_2$, the origin of the translated system will be at
($x_1$ + $x_2$)/2, ($y_1$ + $y_2$)/2, ($z_1$ + $z_2$)/2.  The formula for the hyperboloid, expressed in the Earth-centered system,
becomes

\begin{eqnarray*}
(x_1*\cos(\alpha z_{12}) + x_2*\cos(\alpha z_{12}) +y_1*\cos(\beta z_{12}) + y_2*\cos(\beta z_{12}) + z_1*\cos(\gamma z_{12}) + \\
z_2*\cos(\gamma z_{12}) - 2*x*\cos(\alpha z_{12}) - 2*y*\cos(\beta z_{12}) - 2*z*\cos(\gamma z_{12}))^2/(4*a_{12}^2) \\
-(x_1*\cos(\alpha x_{12}) + x_2*\cos(\alpha x_{12}) + y_1*\cos(\beta x_{12}) + y_2*\cos(\beta x_{12}) + z_1*\cos(\gamma x_{12}) + \\
z_2*\cos(\gamma x_{12}) - 2*x*\cos(\alpha x_{12}) -2*y*\cos(\beta x_{12}) -2*z*\cos(\gamma x_{12}))^2/(4*b_{12}^2) \\
-(x_1*\cos(\alpha y_{12}) + x_2*\cos(\alpha y_{12}) +y_1*\cos(\beta y_{12}) + y_2*\cos(\beta y_{12}) +z_1*\cos(\gamma y_{12}) + \\
z_2*\cos(\gamma y_{12}) -2*x*\cos(\alpha y_{12}) - 2*y*\cos(\beta y_{12}) - 2*z*\cos(\gamma y_{12}))^2/(4*b_{12}^2) -1 = 0
\end{eqnarray*}

where $a_{12}$ and $b_{12}$ refer to the hyperboloid for spacecraft 1 and 2.  A similar equation describes the hyperboloid for
spacecraft 1 and 3.  Although a third equation can be derived for spacecraft 2 and 3, it is not independent of the other two,
because it is constrained by the condition $\delta t_{12}$ + $\delta t_{13}$ + $\delta t_{32}$ = 0.

\begin{figure}
\begin{center}
\includegraphics[width=\textwidth]{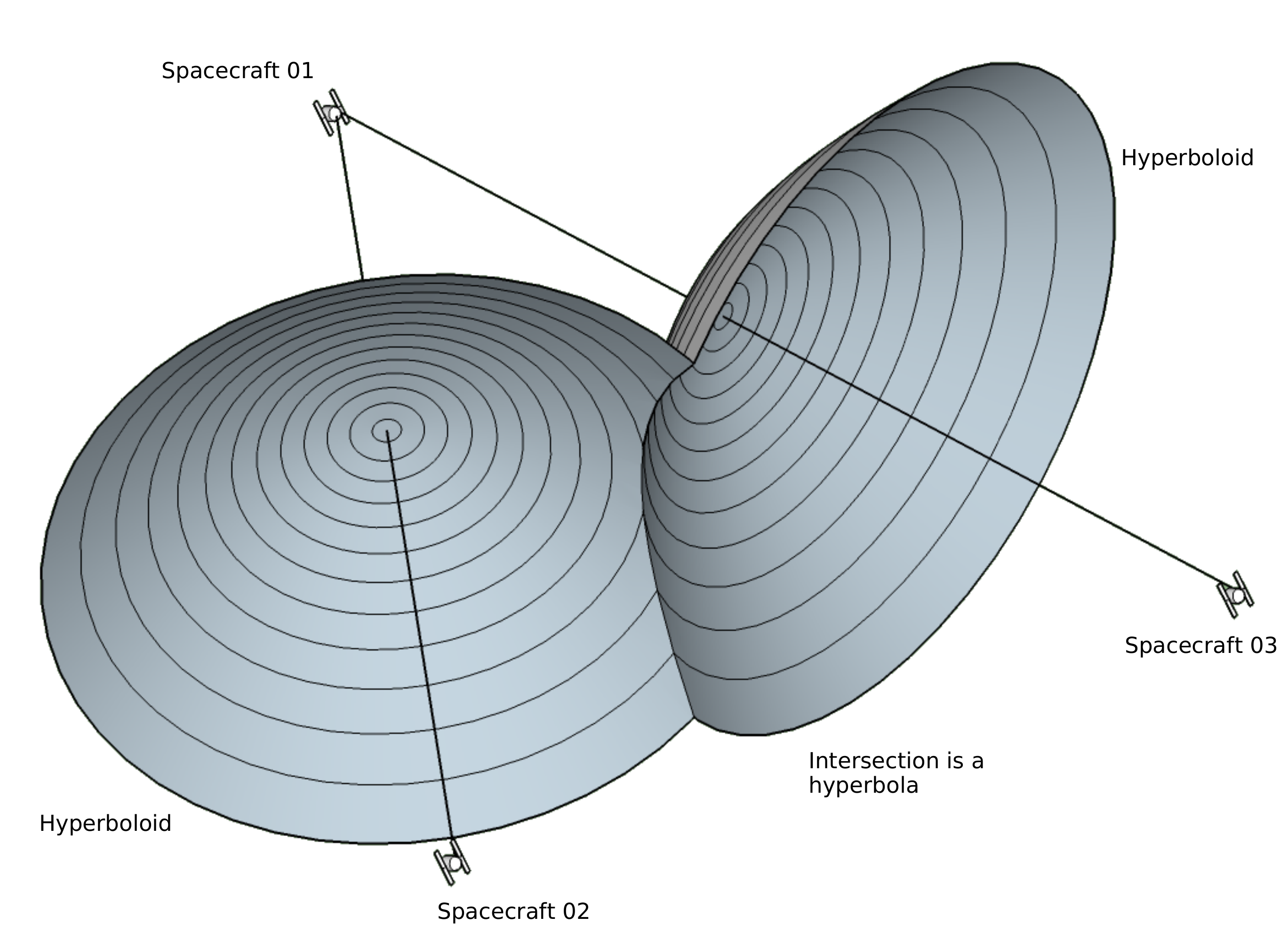}
\caption{The intersection of two hyperboloids. 
With detections with three spacecraft, each spacecraft pair constrain the location of the source
to a surface of a hyperboloid. 
The intersection of the two hyperboloids is a simple hyperbola contained in a plane. The point
on this hyperbola which is closest to Earth gives the lower limit to the source distance.
\label{fig:figure3}}
\end{center}
\end{figure}

The locus of points describing the intersection of two hyperboloids is a simple hyperbola, contained in a plane. This is shown in Figure~\ref{fig:figure3}. This hyperbola
contains all the points satisfying the time delays for the two spacecraft pairs, $\delta t_{12}$ and $\delta t_{13}$, when the GRB
distance is allowed to vary.  The two branches of the hyperbola intersect the celestial sphere at two points; if the distance is taken to be infinite,
the two points are the possible source locations.  It follows that a GRB observed by three, and only three, widely separated non-imaging spacecraft, cannot be unambiguously proven
to originate at a local distance; on the other hand, in the case where the hyperbola degenerates to a single point, that point must be at an infinite
distance, and a local origin can be ruled out.  None of the bursts in this sample were in this category.

In the simplest case, we have one spacecraft near Earth, and two spacecraft in interplanetary space.  So

\begin{eqnarray}
\sqrt{(x-x_1)^2 + (y-y_1)^2 + (z-z_1)^2} - \sqrt{(x-x_2)^2 + (y-y_2)^2 + (z-z_2)^2} - c*\delta t_{12} =0 \\
\sqrt{(x-x_1)^2 + (y-y_1)^2 + (z-z_1)^2} - \sqrt{(x-x_3)^2 + (y-y_3)^2 + (z-z_3)^2} - c*\delta t_{13} =0
\end{eqnarray}

The lower limit to
the source distance is the point on the hyperbola ($x,y,z$) which is closest to Earth.  This can be found by solving for the minimum
of the expression $\sqrt{x^2 + y^2 + z^2}$ (the Earth distance) subject to the constraints imposed by equations A1 and A2.  In practice there are uncertainties
associated with $\delta t_{12}$ and $\delta t_{13}$, and we have used the most probable values to derive the lower limits.  Since $x$, $y$, and $z$ vary along
the hyperbola, the apparent arrival direction for an observer at Earth depends on the assumed distance; if the source distance is assumed to be infinite, the derived right ascension
and declination will not be correct if the source is actually local.  For example, if GRB 140807 were at its minimum allowable distance ($1.46\times 10^{18}$ cm, or $9.8\times 10^{4}$
AU), the angle between its true coordinates and the coordinates for an infinitely distant source would be $0.04\arcdeg$.  But for GRB 101129, whose minimum allowable distance is
only $3.89\times 10^{13}$ cm, or 2.6 AU, the angle would be 54.2\arcdeg.

\begin{figure}
\begin{center}
\includegraphics[width=\textwidth]{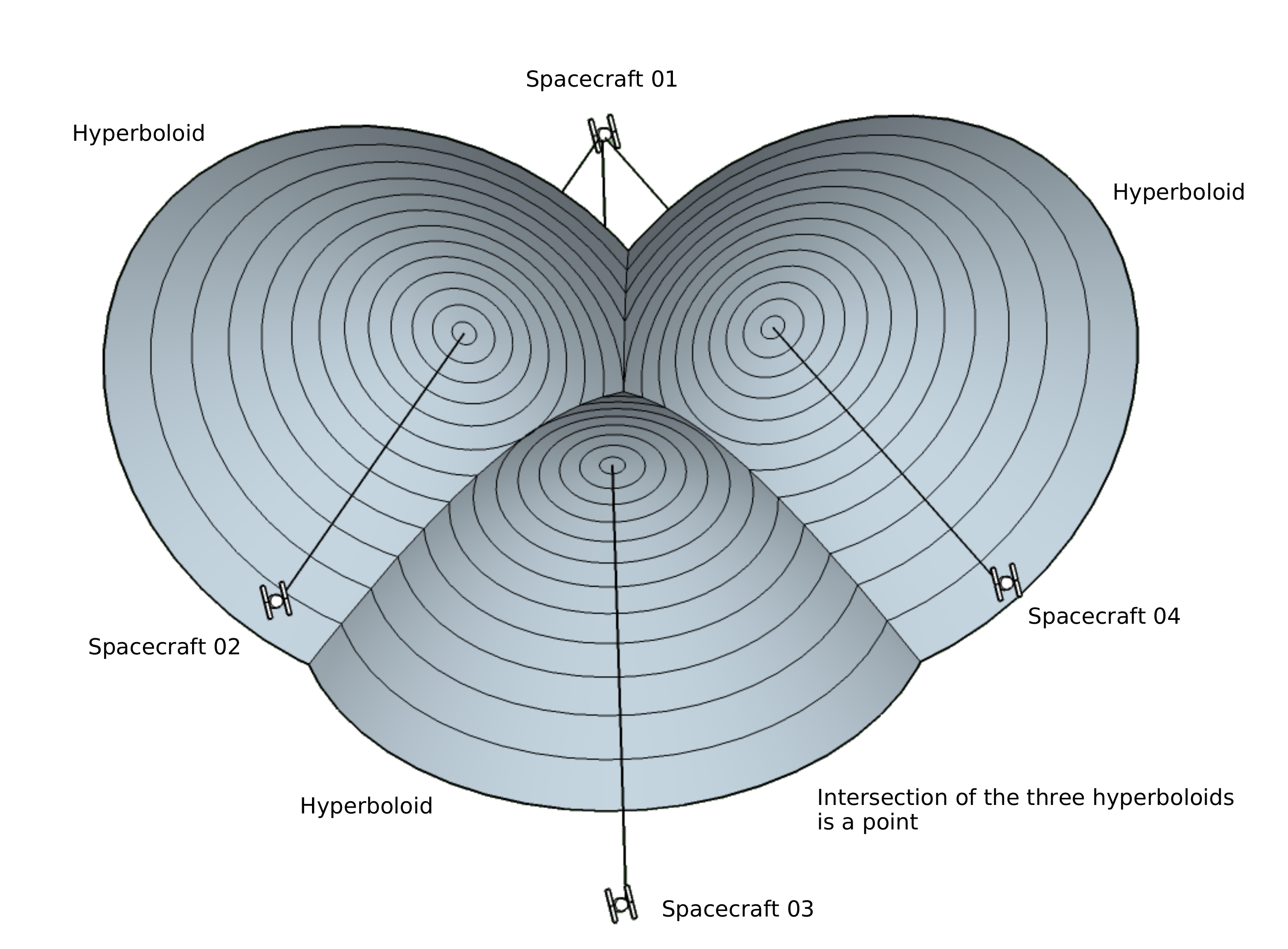}
\caption{The intersection of three hyperboloids. 
With detections with four widely separated spacecraft, each spacecraft pair constrain the location of the source to a surface of a hyperboloid. The intersection of the three hyperboloids is a point in space
as illustrated in the figure. In this case one can determine the source distance purely by
timing measurements.
\label{fig:figure4}}
\end{center}
\end{figure}

\begin{figure}
\begin{center}
\includegraphics[width=0.8\textwidth]{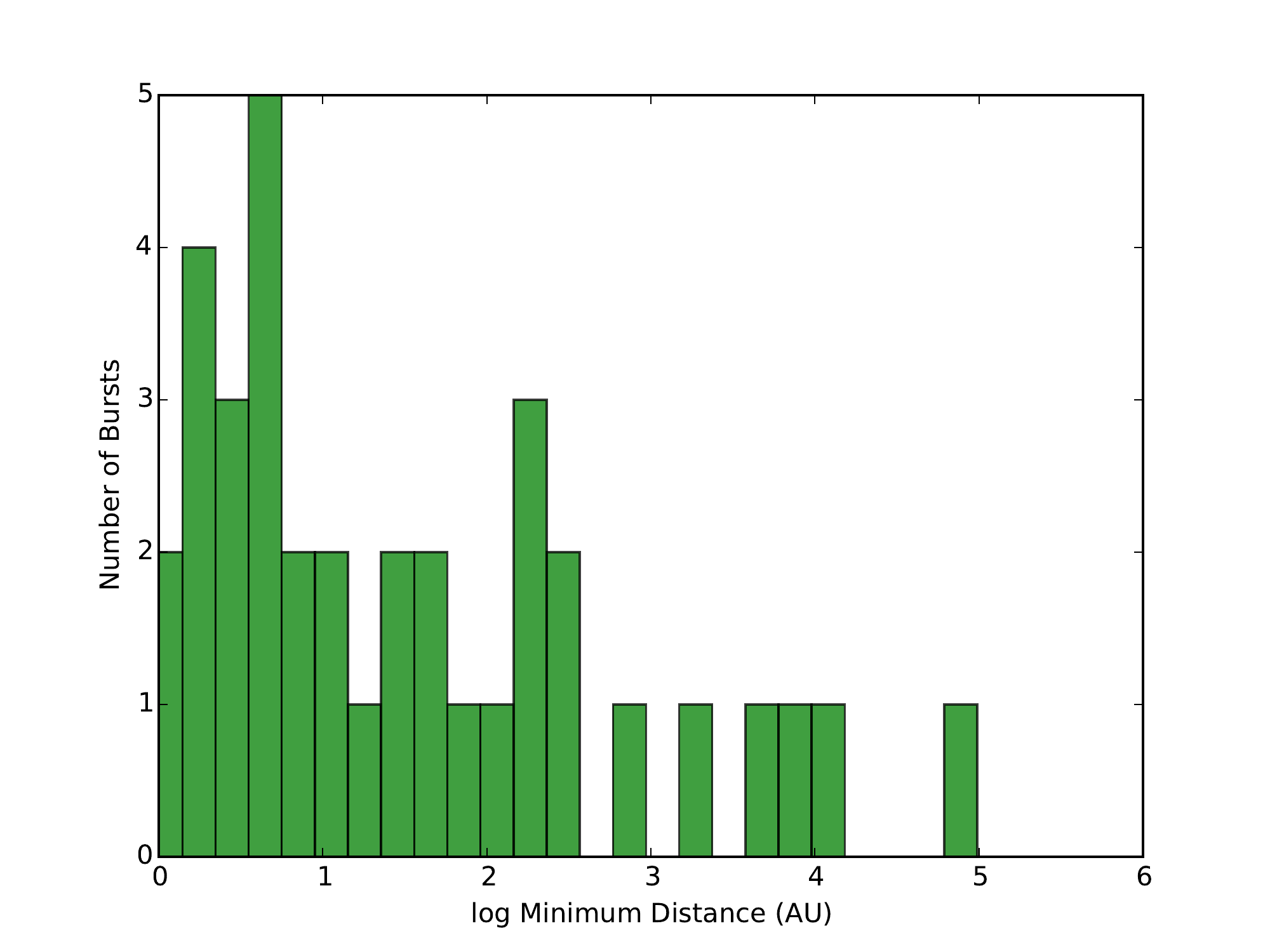}
\caption{Histogram of the minimum distances to the PBH burst candidates in the IPN sample.
\label{fig:pbh_min_dist_histo}}
\end{center}
\end{figure}

\begin{figure}
\begin{center}
\includegraphics[width=0.99\textwidth]{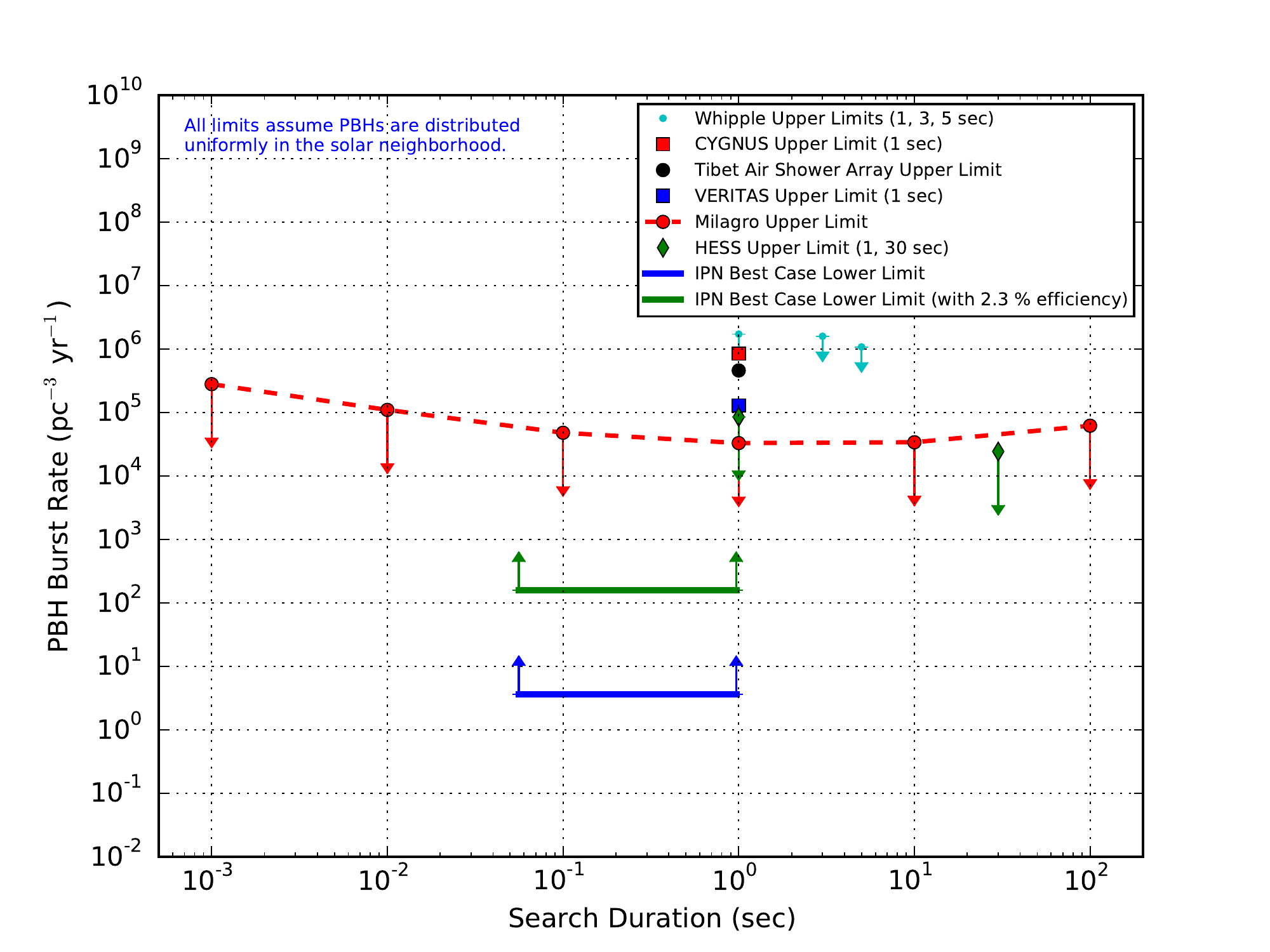}
\caption{IPN PBH burst rate lower limit estimates assuming all the candidates are real PBH bursts. The horizontal green line gives the IPN PBH burst rate lower limit considering the selection efficiency of 2.3\%. The blue horizontal line shows the lower limit if we assume 100\% selection efficiency. Published PBH burst rate upper limits for various other burst search experiments are shown for comparison~\citep{alex1993, amenomori1995, linton2006, veritas2012, glicenstein2013, abdo2015}.
\label{fig:pbh-limit-final}}
\end{center}
\end{figure}

\begin{figure}
\begin{center}
\includegraphics[width=0.9\textwidth]{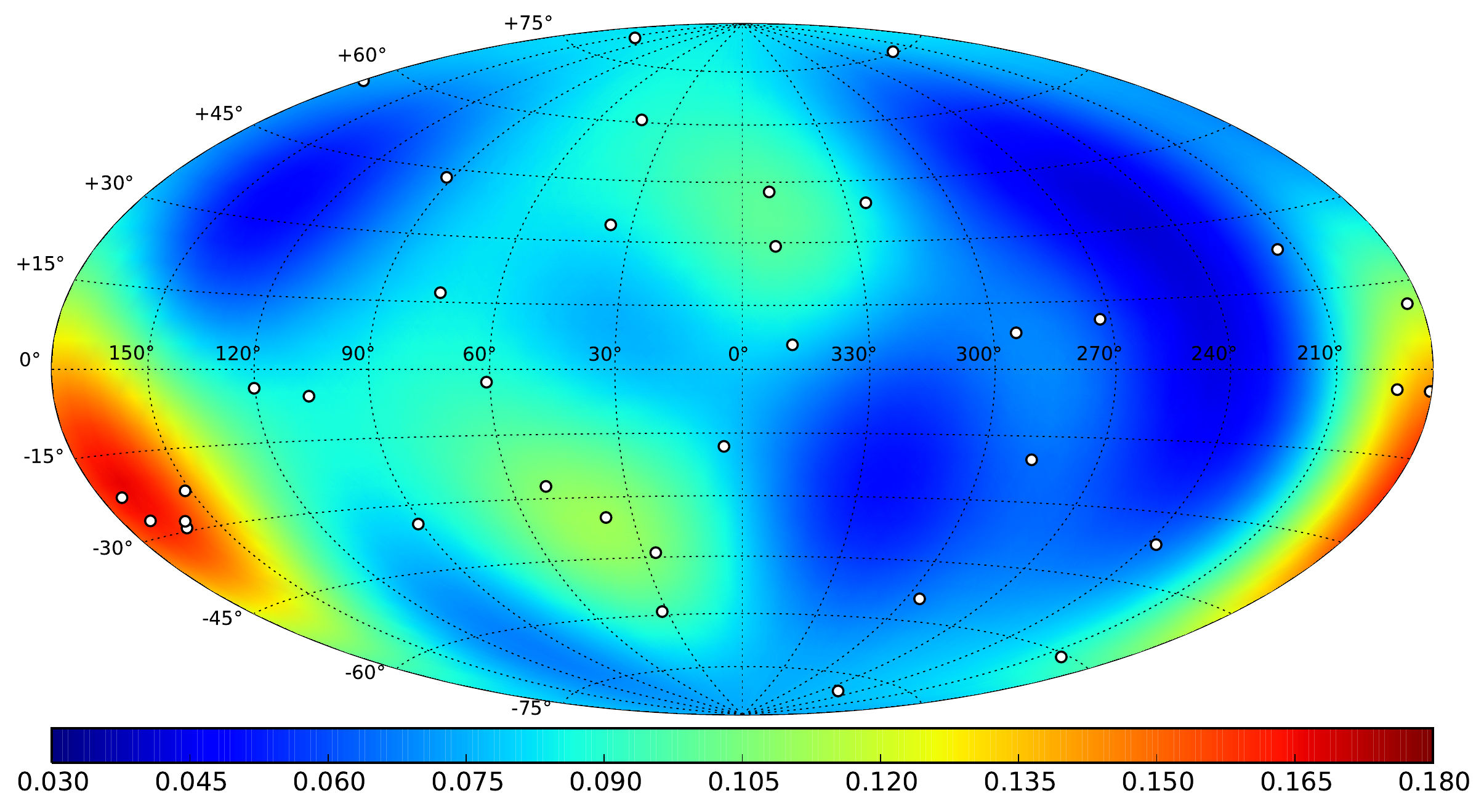}
\caption{GRB density map in Galactic coordinates for the PBH burst candidates
assuming minimum distances given in Table~\ref{distances}.
This map is normalized to represent a probability density function (PDF) 
that integrates to 1 over the entire sphere. 
The smoothing parameter is taken to be 25 degrees.
The circles indicate the locations of the individual bursts.
The maximum and the minimum density values in this map are 0.166 and 0.041,
respectively. The probability of generating this density contrast by chance in the case when the true sky distribution is uniform, is $\sim$0.2 estimated using a Monte Carlo simulation ~\citep{ukwatta2015}. Thus the density structures
seen in the map are consistent with a uniform distribution.
\label{sky_map_pbh_candidate_min}}
\end{center}
\end{figure}

\begin{figure}
\begin{center}
\includegraphics[width=0.9\textwidth]{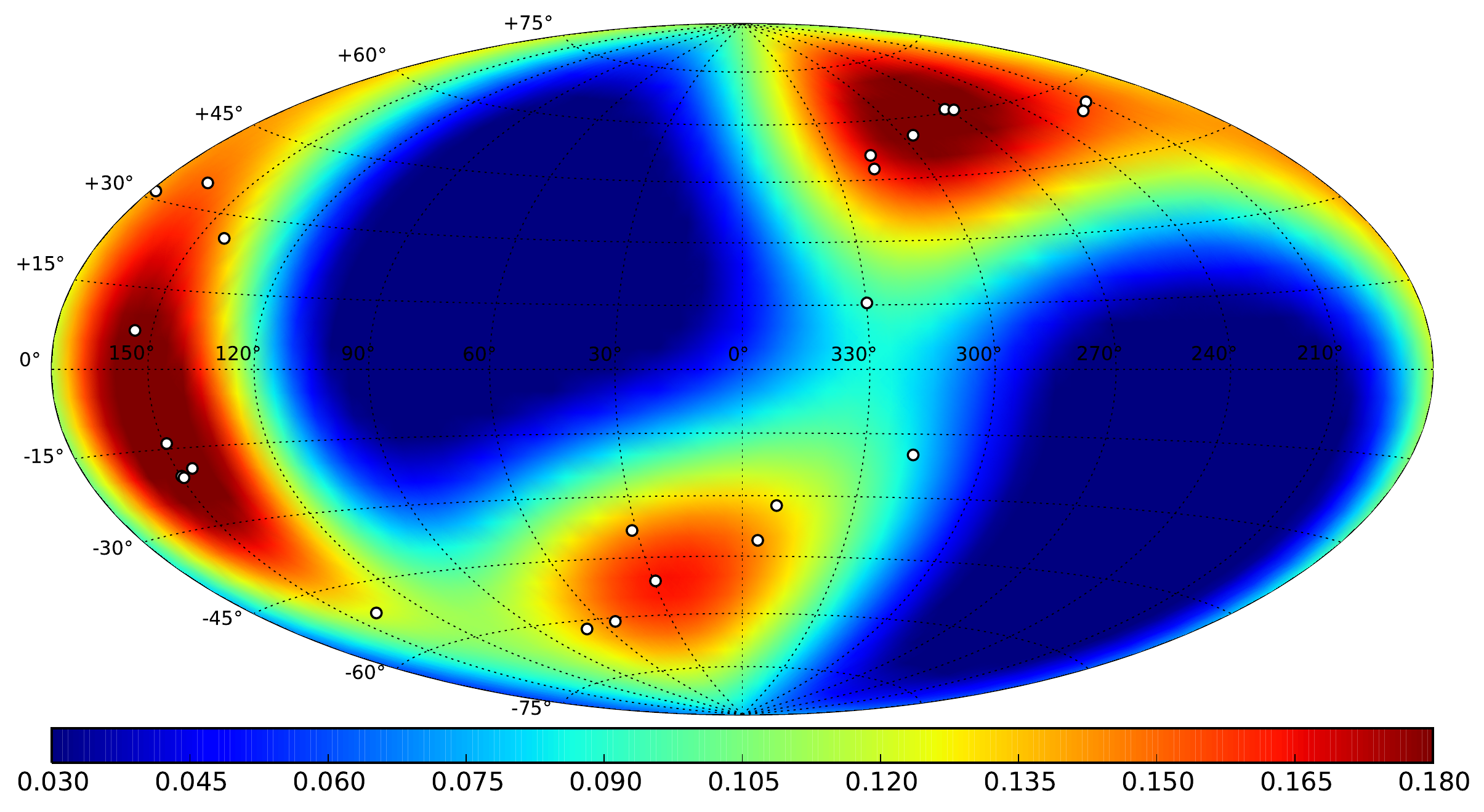}
\caption{GRB density map in Galactic coordinates for the PBH burst candidates
assuming constant distances of 10 parsecs to sources. The map has only 24 candidates with four spacecraft detections. Remaining 8 bursts have only
three spacecraft detections, so they don't have a single localization.
This map is normalized to represent a probability density function (PDF) 
that integrates to 1 over the entire sphere. 
The smoothing parameter is taken to be 25 degrees.
The circles indicate the locations of the individual bursts.
The maximum and the minimum density values in this map are 0.202 and 0.002,
respectively. The probability of generating this density contrast by chance in the case when the true sky distribution is uniform, is $\sim$0.1 estimated using a Monte Carlo simulation ~\citep{ukwatta2015}. Thus the density structures
seen in the map are consistent with a uniform distribution.
\label{sky_map_pbh_candidate_10pc}}
\end{center}
\end{figure}

\begin{figure}
\begin{center}
\includegraphics[width=1.0\textwidth]{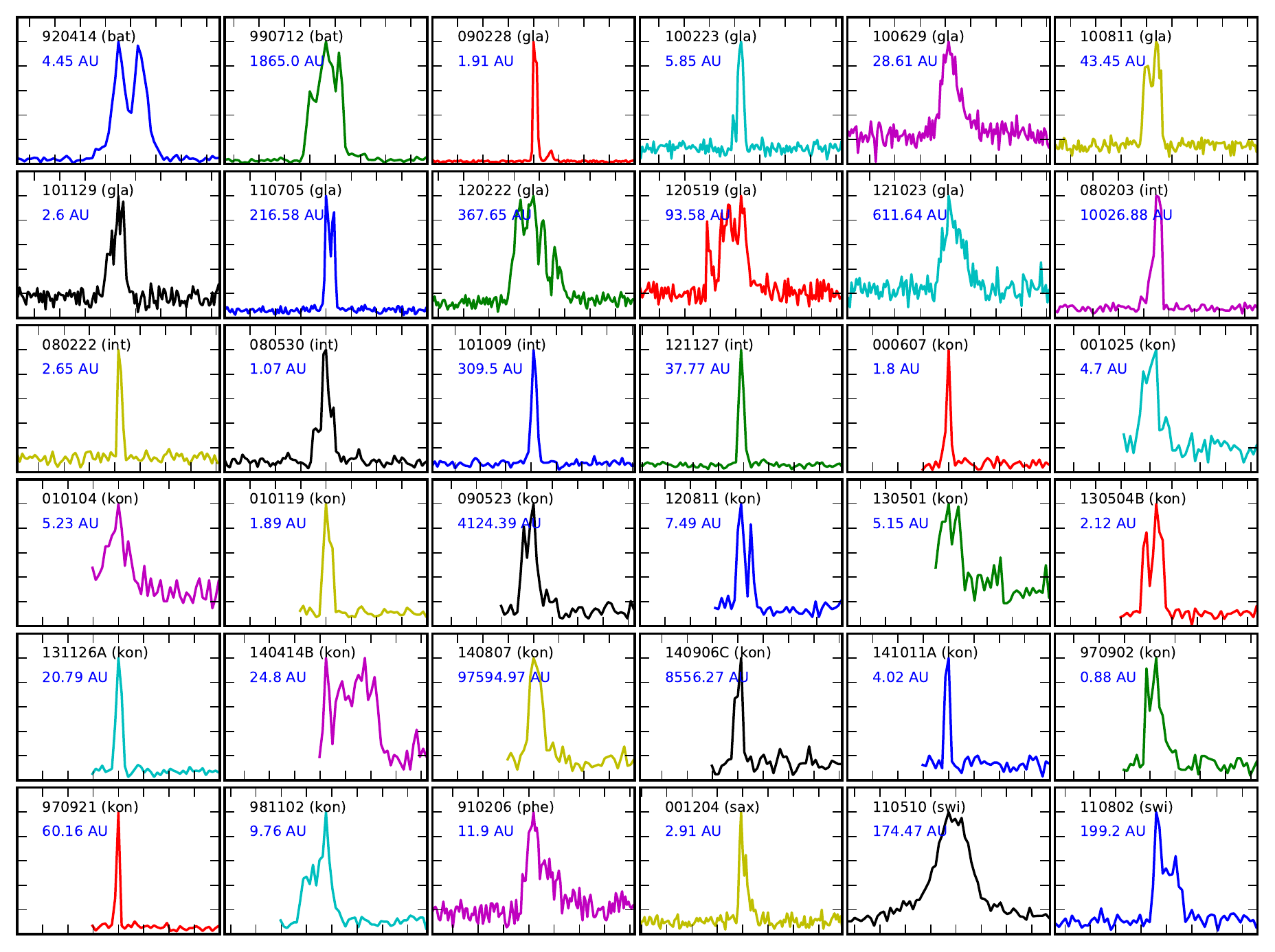}
\caption{Normalized light curves of all the PBH burst candidates in the IPN sample. Black labels show the burst name with
parenthesis showing the instrument. The blue labels give the minimum distance to the
bursts based on our analysis. Each light curve shows a time range of 4 seconds centered on
the brightest peak.
\label{lc_grid}}
\end{center}
\end{figure}

In a number of cases, a fourth non-coplanar experiment, in this case \it Konus-Wind \rm, at up to 7 light-seconds from Earth, can be used to constrain
the lower limits further.  Adding the constraint

\begin{eqnarray*}
\sqrt{(x-x_1)^2 + (y-y_1)^2 + (z-z_1)^2} - \sqrt{(x-x_4)^2 + (y-y_4)^2 + (z-z_4)^2} - c*\delta t_{14} =0
\end{eqnarray*}

eliminates one branch of the hyperbola and part of the second branch, leading to a larger distance lower limit.  Thus, in
the case of four widely separated spacecraft, it is in principle possible to rule out an infinite distance and prove that the origin is local as illustrated in Figure~\ref{fig:figure4}.
However, this was not the case for any of the bursts in this study; they are all consistent with both local and infinite distances.

Note that this method does not depend on the properties of the GRB itself, such as duration or intensity; the lower
limit is determined by the IPN configuration (through the spacecraft coordinates) and the direction of the burst (through
the time delays).  Thus, for example, it can be applied equally to long- and short-duration bursts, and a sample of
long-duration events would yield a distribution of lower limits which was comparable to a sample of short-duration events.

One special case should be noted here.  With just two widely separated spacecraft, if one has precise imaging capability, the problem reduces to finding the intersection of a hyperboloid and the vector defined by the precise localization from the imager. In principle, a local origin can be demonstrated, or a distance lower limit can be obtained.  We have studied approximately two hundred GRBs which are in this category, and analysis of the results is ongoing.

\clearpage

\clearpage

\begin{deluxetable}{ccccc}

\setlength{\tabcolsep}{.02in}
\rotate
\tabletypesize{\scriptsize}

\tablecaption{Comparison of methods used to detect PBH evaporations. \label{comparison}}
\tablewidth{0pt}
\tablehead{
\colhead{Method}  & \colhead{Burst duration, s} & \colhead{Energy or frequency} & \colhead{Rate upper limits, pc$^{-3}$ y$^{-1}$} & \colhead{References}
}
\startdata
Atmospheric Cherenkov	&  $10^{-7}-0.1$      & 200 Mev-10 TeV                                 & $0.04-8.7\times 10^{5}$         & 1               \\
Air Shower              &  $10^{-6}-0.1$      & $5\times 10^{12}-\gtrsim 5\times 10^{13}$ eV   & $2.7\times 10^3-6\times 10^3$   & 2               \\
Radio Pulse             &  $<3\times 10^{-3}$ & 430-1374 MHz                                   & $2\times 10^{-9}$               & 3               \\
Spark Chamber           &  $10^{-6}$          & 20 MeV-1 GeV                                   & $5\times 10^{-2}$               & 4               \\
Spatial Distribution    &  $<0.1$             & 15 keV-10 MeV                                  & \nodata                         & 5               \\
\enddata

\tablerefs{
(1) \citet{pw1,pw2,pw3,abdo2015}; (2) \citet{bh1,fe1}; (3) \citet{pt1,ke1}; (4)  \citet{fi1}; (5) \citet{c1,c2,c3,c4,c5,c6,c7}
}

\end{deluxetable}

\clearpage

\begin{deluxetable}{ccccccccccccccc}

\setlength{\tabcolsep}{.02in}
\rotate
\tabletypesize{\scriptsize}

\tablecaption{Distance lower limits for IPN gamma-ray bursts. \label{distances}}
\tablewidth{0pt}
\tablehead{
\colhead{GRB} & \colhead{SOD} & \colhead{Spacecraft\tablenotemark{a}} & \colhead{Dur.,} & \colhead{Fluence}              & \colhead{Energy} & \colhead{Minimum}   & \colhead{Maximum}    & \colhead{Maximum}                               & \colhead{Maximum}                                & \colhead{Counterpart} & \colhead{Refs.} \\
\colhead{}    & \colhead{}    & \colhead{}                            & \colhead{s}     & \colhead{erg cm$^{-2}$}        & \colhead{range,} & \colhead{Distance,} & \colhead{detectable} & \colhead{detectable}                            & \colhead{detectable}                             & \colhead{search?}     & \colhead{}      \\
\colhead{}    & \colhead{}    & \colhead{}                            & \colhead{}      & \colhead{}                     & \colhead{keV}    & \colhead{cm.}       & \colhead{dist., cm.} & \colhead{dist., $({\eta_{\gamma D}})^{0.5}$ cm.} & \colhead{dist., $({\eta'_{\gamma D}})^{0.5}$ cm.} & \colhead{}            & \colhead{}      \\
\colhead{}    & \colhead{}    & \colhead{}                            & \colhead{}      & \colhead{}                     & \colhead{}       & \colhead{}          & \colhead{}           & \colhead{(SEM Model)}                           & \colhead{(HM Model)}                             & \colhead{}            & \colhead{}
}
\startdata
910206    &  31529  & Uly,PVO,Phe        & 0.7   & $1.1 \times 10^{-5}$   & 100--100000 & $1.8 \times 10^{14}$   & $8.5 \times 10^{18}$   & $8.8 \times 10^{16}$   & $2.6 \times 10^{19}$   & No         & 1,6               \\ 
920414    &  84162  & Uly,BAT,PVO        & 0.96  & $5.8 \times 10^{-6}$   & 20 --2000   & $6.7 \times 10^{13}$   & $1.2 \times 10^{19}$   & $1.3 \times 10^{17}$   & $3.7 \times 10^{19}$   & No         & 2,3,6             \\ 
970902    &  27561  & Uly,Kon,NEAR       & 0.44  & $4.2 \times 10^{-6}$   & 10 --10000  & $1.3 \times 10^{13}$   & $1.4 \times 10^{19}$   & $1.3 \times 10^{17}$   & $4.3 \times 10^{19}$   & No         & 4,5               \\ 
970921    &  83828  & Uly,Kon,NEAR       & 0.06  & $2.9 \times 10^{-6}$   & 10 --10000  & $9.0 \times 10^{14}$   & $1.6 \times 10^{19}$   & $1.1 \times 10^{17}$   & $5.1 \times 10^{19}$   & No         & 4,5               \\ 
981102    &  28554  & Uly,Kon,NEAR       & 0.57  & $1.1 \times 10^{-5}$   & 10 --10000  & $1.5 \times 10^{14}$   & $8.5 \times 10^{18}$   & $8.5 \times 10^{16}$   & $2.7 \times 10^{19}$   & No         & 5,13              \\ 
990712    &  27919  & Uly,BAT,Kon,NEAR   & 0.62  & $2.1 \times 10^{-5}$   & 10 --10000  & $2.8 \times 10^{16}$   & $1.3 \times 10^{19}$   & $6.2 \times 10^{16}$   & $1.9 \times 10^{19}$   & No         & 5,8               \\ 
000607    &  08689  & Uly,Kon,NEAR       & 0.09  & $4.6 \times 10^{-6}$   & 10 --10000  & $2.7 \times 10^{13}$   & $1.4 \times 10^{19}$   & $9.6 \times 10^{16}$   & $4.1 \times 10^{19}$   & 14,15      & 5,13              \\ 
001025    &  71369  & Uly,Kon,NEAR       & 0.48  & $4.9 \times 10^{-6}$   & 10 --10000  & $7.0 \times 10^{13}$   & $1.4 \times 10^{19}$   & $1.2 \times 10^{17}$   & $4.0 \times 10^{19}$   & 15,16      & 5,13              \\ 
001204    &  28869  & Uly,Kon,SAX,NEAR   & 0.27  & $1.1 \times 10^{-6}$   & 10 --10000  & $4.4 \times 10^{13}$   & $3.0 \times 10^{19}$   & $2.4 \times 10^{17}$   & $8.4 \times 10^{19}$   & 15,17,18   & 5,7,13            \\ 
010104    &  62490  & Uly,Kon,SAX,NEAR   & 0.76  & $6.6 \times 10^{-7}$   & 10 --10000  & $7.8 \times 10^{13}$   & $3.6 \times 10^{19}$   & $3.6 \times 10^{17}$   & $1.1 \times 10^{20}$   & No         & 5,7,13            \\ 
010119    &  37177  & Uly,Kon,NEAR       & 0.18  & $2.3 \times 10^{-6}$   & 10 --10000  & $2.8 \times 10^{13}$   & $1.9 \times 10^{19}$   & $1.5 \times 10^{17}$   & $5.8 \times 10^{19}$   & No         & 5,7,13            \\ 
080203    &  08456  & Ody,Kon,INT,MES    & 0.4   & $9.3 \times 10^{-6}$   & 10 --10000  & $1.5 \times 10^{17}$   & $9.8 \times 10^{18}$   & $8.7 \times 10^{16}$   & $2.9 \times 10^{19}$   & No         & 5,13              \\ 
080222    &  37262  & Ody,Kon,MES        & 0.11  & $1.9 \times 10^{-6}$   & 10 --10000  & $4.0 \times 10^{13}$   & $2.1 \times 10^{19}$   & $1.6 \times 10^{17}$   & $6.4 \times 10^{19}$   & No         & 5,13              \\ 
080530    &  58296  & Ody,INT,MES,Suz    & 0.41  & $1.0 \times 10^{-6}$   & 100--1000   & $1.6 \times 10^{13}$   & $2.8 \times 10^{19}$   & $2.6 \times 10^{17}$   & $8.7 \times 10^{19}$   & No         & 6                 \\ 
090228    &  17600  & Ody,Kon,RHE,MES    & 0.08  & $6.3 \times 10^{-6}$   & 10 --10000  & $2.9 \times 10^{13}$   & $1.1 \times 10^{19}$   & $8.1 \times 10^{16}$   & $3.5 \times 10^{19}$   & No         & 5,6,9,10,19       \\ 
090523    &  34075  & Ody,Kon,MES,Suz    & 0.5   & $1.6 \times 10^{-6}$   & 10 --10000  & $6.2 \times 10^{16}$   & $2.3 \times 10^{19}$   & $2.2 \times 10^{17}$   & $6.9 \times 10^{19}$   & No         & 5,6,13            \\ 
100223    &  09491  & Ody,Kon,MES,Suz    & 0.19  & $1.7 \times 10^{-6}$   & 10 --10000  & $8.8 \times 10^{13}$   & $2.3 \times 10^{19}$   & $1.8 \times 10^{17}$   & $6.7 \times 10^{19}$   & No         & 5,6,9,10          \\ 
100629    &  69243  & Ody,Kon,INT,MES    & 0.46  & $1.1 \times 10^{-6}$   & 10 --10000  & $4.3 \times 10^{14}$   & $2.6 \times 10^{19}$   & $2.6 \times 10^{17}$   & $8.5 \times 10^{19}$   & No         & 5,6,9,10          \\ 
100811    &  09349  & Ody,Kon,INT,MES    & 0.33  & $4.4 \times 10^{-6}$   & 10 --10000  & $6.5 \times 10^{14}$   & $1.6 \times 10^{19}$   & $1.2 \times 10^{17}$   & $4.2 \times 10^{19}$   & No         & 5,6,9,11          \\ 
101009    &  24858  & Ody,Kon,INT,MES    & 0.18  & $1.6 \times 10^{-6}$   & 10 --10000  & $4.6 \times 10^{15}$   & $1.6 \times 10^{19}$   & $1.8 \times 10^{17}$   & $6.9 \times 10^{19}$   & No         & 5,6,13            \\ 
101129    &  56371  & Ody,Kon,INT,MES    & 0.32  & $3.7 \times 10^{-6}$   & 10 --10000  & $3.9 \times 10^{13}$   & $3.0 \times 10^{19}$   & $1.3 \times 10^{17}$   & $4.6 \times 10^{19}$   & No         & 5,9,11            \\ 
110510    &  80844  & Ody,Kon,Swi,MES    & 0.84  & $9.7 \times 10^{-7}$   & 10 --10000  & $2.6 \times 10^{15}$   & $3.1 \times 10^{19}$   & $3.0 \times 10^{17}$   & $8.9 \times 10^{19}$   & No         & 6,13              \\ 
110705    &  13031  & Ody,Kon,INT,MES    & 0.22  & $5.7 \times 10^{-6}$   & 10 --10000  & $3.2 \times 10^{15}$   & $1.7 \times 10^{19}$   & $1.0 \times 10^{17}$   & $3.7 \times 10^{19}$   & No         & 9,11,20,21,22     \\ 
110802    &  55157  & Ody,Kon,INT,MES    & 0.54  & $1.3 \times 10^{-5}$   & 10 --10000  & $3.0 \times 10^{15}$   & $7.8 \times 10^{18}$   & $7.6 \times 10^{16}$   & $2.4 \times 10^{19}$   & No         & 13,23,24          \\ 
120222    &  01776  & Ody,Kon,MES,Fer    & 0.87  & $1.8 \times 10^{-6}$   & 10 --10000  & $5.5 \times 10^{15}$   & $2.1 \times 10^{19}$   & $2.3 \times 10^{17}$   & $6.6 \times 10^{19}$   & No         & 9,11              \\ 
120519    &  62294  & Ody,Kon,MES,Fer    & 0.91  & $3.6 \times 10^{-6}$   & 10 --10000  & $1.4 \times 10^{15}$   & $1.8 \times 10^{19}$   & $1.6 \times 10^{17}$   & $4.6 \times 10^{19}$   & No         & 9,11,25,26,27     \\ 
120811    &  01230  & Ody,Kon,MES,Fer    & 0.31  & $4.3 \times 10^{-6}$   & 10 --10000  & $1.1 \times 10^{14}$   & $1.9 \times 10^{19}$   & $1.2 \times 10^{17}$   & $4.2 \times 10^{19}$   & No         & 9,11,28,29,30,31  \\ 
121023    &  27857  & Ody,MES,Fer        & 0.51  & $7.7 \times 10^{-7}$   & 10 --1000   & $9.2 \times 10^{15}$   & $3.2 \times 10^{19}$   & $3.1 \times 10^{17}$   & $1.0 \times 10^{20}$   & No         & 9,11              \\ 
121127    &  78960  & Ody,Kon,INT,MES    & 0.53  & $2.8 \times 10^{-6}$   & 10 --10000  & $5.6 \times 10^{14}$   & $4.1 \times 10^{19}$   & $1.7 \times 10^{17}$   & $5.3 \times 10^{19}$   & No         & 11,12,32,33,34    \\ 
130501    &  00831  & Ody,Kon,MES,Suz    & 0.41  & $2.0 \times 10^{-6}$   & 10 --10000  & $7.7 \times 10^{13}$   & $2.0 \times 10^{19}$   & $1.9 \times 10^{17}$   & $6.2 \times 10^{19}$   & No         & 13                \\ 
130504B   &  27123  & Ody,Kon,MES        & 0.36  & $8.8 \times 10^{-6}$   & 10 --10000  & $3.2 \times 10^{13}$   & $9.2 \times 10^{18}$   & $8.8 \times 10^{16}$   & $3.0 \times 10^{19}$   & No         & 35,36,37,38       \\ 
131126A   &  14050  & Ody,Kon,MES,Fer    & 0.12  & $2.3 \times 10^{-6}$   & 10 --10000  & $3.1 \times 10^{14}$   & $2.0 \times 10^{19}$   & $1.4 \times 10^{17}$   & $5.8 \times 10^{19}$   & 39         & 40,41,42          \\ 
140414B   &  80735  & Ody,Kon,INT,MES    & 0.97  & $2.3 \times 10^{-6}$   & 10 --10000  & $3.7 \times 10^{14}$   & $2.0 \times 10^{19}$   & $2.0 \times 10^{17}$   & $5.9 \times 10^{19}$   & No         & 13                \\ 
140807    &  43173  & Ody,Kon,MES,Fer    & 0.57  & $2.6 \times 10^{-6}$   & 10 --10000  & $1.5 \times 10^{18}$   & $1.8 \times 10^{19}$   & $1.8 \times 10^{17}$   & $5.5 \times 10^{19}$   & 43         & 13                \\ 
140906C   &  85869  & Ody,Kon,INT,MES    & 0.13  & $2.3 \times 10^{-6}$   & 10 --10000  & $1.3 \times 10^{17}$   & $1.9 \times 10^{19}$   & $1.5 \times 10^{17}$   & $5.8 \times 10^{19}$   & No         & 44,45             \\ 
141011A   &  24380  & Ody,Kon,MES,Fer    & 0.06  & $1.4 \times 10^{-6}$   & 10 --10000  & $6.0 \times 10^{13}$   & $2.5 \times 10^{19}$   & $1.6 \times 10^{17}$   & $7.3 \times 10^{19}$   & No         & 46,47,48          \\ 
\enddata

\tablenotetext{\,\,\,\,\,\,\,\,\,\,\,\,\,\,\,\,\,\,\,\,a}{
Fer: \it Fermi \rm,
INT: \it International Gamma-Ray Laboratory\rm,
Kon: \it Konus-Wind\rm, \\
MES: \it Mercury Surface, Space Environment, Geochemistry, and Ranging \rm mission,
NEAR: \it Near Earth Asteroid Rendezvous \rm mission,
Ody: \it Mars Odyssey\rm,
Phe: \it Phebus\rm,
PVO: \it Pioneer Venus Orbiter \rm, \\
RHE: \it Ramaty High Energy Solar Spectroscopic Imager\rm,
SAX: \it Satellite per Astronomia X (BeppoSAX) \rm,
Suz: \it Suzaku\rm,
Swi: \it Swift \rm (burst was outside the coded field of view of the BAT, and not localized by it),
Uly: \it Ulysses \rm
 \\
}
\tablerefs{
(1) \citet{t1}; (2) \citet{g1}; (3) \citet{pa1}; (4) \url{http://www.ioffe.ru/LEA/shortGRBs/Catalog/};
(5) \citet{pa2}; (6) \url{http://ssl.berkeley.edu/ipn3/index.html}; (7) \citet{f1};
(8) \url{http://www.batse.msfc.nasa.gov/batse/grb/catalog/current}; (9) \citet{g2};
(10) \citet{pa3}; (11) \citet{v1}; (12) \citet{g2}; (13) \citet{s2}; (14)\citet{m2};
(15) \citet{h3}; (16) \citet{p3}; (17) \citet{p4}; (18) \citet{v2}; (19) \citet{g4};
(20) \citet{g5}; (21) \citet{g6}; (22) \citet{y1}; (23) \citet{h4}; (24) \citet{g7};
(25) \citet{g9}; (26) \citet{g10}; (27) \citet{s3};
(28) \citet{g11}; (29) \citet{g12}; (30) \citet{g13}; (31) \citet{x1}; (32) \citet{g14};
(33) \citet{g15}; (34) \citet{i1}; (35) \citet{v2}; (36) \citet{g16}; (37) \citet{g17};
(38) \citet{y2}; (39) \citet{si1}; (40) \citet{g18}; (41) \citet{g19}; (42) \citet{pe1};
(43) \citet{si2}; (44) \citet{g20} ; (45) \citet{g21}; (46) \citet{v3}; (47) \citet{g22};
(48) \citet{g23}
}

\end{deluxetable}

\begin{deluxetable}{ccccc}

\tabletypesize{\scriptsize}
\tablecaption{Localizations of IPN gamma-ray bursts assuming infinite distances. Some bursts have two possible
error boxes; GRB080203 has an eight-cornered error box.
\label{localizations}}
\tablewidth{0pt}
\tablehead{
\colhead{GRB} & \colhead{$\alpha$} &\colhead{$\delta$}  \\
}
\startdata

910206	&  &  \\
Center  &   87.5650   &  17.3372   \\
Corners &   86.7457   &  17.8617   \\
        &   88.4371   &  16.5674   \\
        &   86.6713   &  18.2592   \\
        &   88.3683   &  16.9025   \\
OR      &  & \\
Center  &   86.9621   &  36.1678   \\
Corners &   86.0419   &  35.5941   \\
        &   87.9516   &  36.9783   \\
        &   85.9902   &  35.1926   \\
        &   87.8875   &  36.6407   \\
\hline
920414  &  &    \\
Center  &   90.5234  &  -76.3115   \\
Corners &   90.6025  &  -76.3287   \\
        &   90.3189  &  -76.2150   \\
        &   90.7309  &  -76.4079   \\
        &   90.4446  &  -76.2943   \\
\hline
970902  &  &      \\
Center  &  351.0259 & +7.3294 \\
Corners &  351.0744 & +7.3080 \\
        &  351.1009 & +7.3100 \\
        &  350.9774 & +7.3507 \\
        &  350.9509 & +7.3486 \\
\hline
970921  &  &      \\
Center  &  235.8826 & -24.8310 \\
Corners &  236.0899 & -24.3906 \\
        & 236.2034  & -24.1627 \\
        & 235.6864  & -25.2436  \\
        & 235.5867  & -25.4367  \\
\hline

981102  &  &    \\
Center  & 277.0514  &  -48.3354 \\
Corners & 277.0532  &   -48.3258 \\
        & 277.0159  &   -48.2744 \\
        & 277.0496  &   -48.3449 \\
        & 277.0870  &   -48.3962 \\
\hline
\tablebreak

990712  &  &     \\
Center  & 123.4627  &  +6.6755 \\
Corners & 123.8544  &   +7.2952 \\
        & 123.2666  &     +6.3351 \\
        & 123.2036  &     +6.2494 \\
        & 123.7301  &     +7.1201 \\
OR      &           &             \\
Center  & 125.3550  &     +9.4408 \\
Corners & 124.9180  &     +8.8500 \\
        & 125.6024  &     +9.7478 \\
        & 125.6597  &     +9.8378 \\
        & 125.0364  &     +9.0294 \\
\hline
000607  &  &     \\
Center  & 38.4971  &  +17.1419 \\
Corners & 38.4656  &   +17.1016 \\
        & 38.5472  &   +17.2317 \\
        & 38.5287  &   +17.1823 \\
        & 38.4470  &   +17.0523 \\
\hline
001025  &  &    \\
Center   & 275.3488  &  -5.1067 \\
Corners  & 275.4272  &    -5.1411 \\
         & 275.1851  &    -4.9912 \\
         & 275.2707  &    -5.0723 \\
         & 275.5140  &    -5.2229 \\
\hline
001204  &  &     \\
Center  &  40.2997  &  +12.8817 \\
Corners &  40.2796  &   +12.8928 \\
        &  40.3478  &   +12.8979 \\
        &  40.3199  &   +12.8707 \\
        &  40.2516  &   +12.8656 \\
\hline
010104  &  &    \\
Center  &  317.3689  & +63.5116 \\
Corners &  317.6201  &  +63.5227 \\
        &  317.5851  &  +63.4776 \\
        &  317.1183  &  +63.5004 \\
        &  317.1523  &  +63.5455 \\
\hline
010119  &  &     \\
Center  &  283.4446  &  +11.9964 \\
Corners &  283.4892  &   +12.0007 \\
        &  283.4159  &   +11.9832 \\
        &  283.4000  &   +11.9921 \\
        &  283.4734  &   +12.0096 \\
\hline
\tablebreak

080203  &  &     \\
Center  &  48.2086  &  +24.3289 \\
Corners &  48.1704  &  +24.3216 \\
        &  48.2401  &   +24.0161 \\
        & 49.1541  &   +20.8021  \\
        & 48.9850  &   +21.5848 \\
        & 48.2469  &   +24.3362 \\
        & 47.7229  &   +27.1306 \\
        & 47.5349  &   +27.9115 \\
        & 48.1002  &   +24.6272 \\
\hline
080222  &  &     \\
Center  &  68.7192 &  +56.9390 \\
Corners &  69.2723 &   +57.0738 \\
        &  68.1039 &   +56.6937 \\
        &  68.1686 &   +56.7979 \\
        &  69.3407 &   +57.1754 \\
\hline
080530  &  &     \\
Center  &   5.3443 &   28.2246  \\
Corners &   5.2701 &   28.3257  \\
        &   5.4170 &   28.1570  \\
        &   5.2717 &   28.2920  \\
        &   5.4186 &   28.1232  \\
\hline
090228  &  &     \\
Center  &  98.7014  &  -28.7863 \\
Corners &  98.7827  &   -28.8531 \\
        &  98.5307  &   -28.6112 \\
        &  98.5698  &   -28.6709 \\
        &  98.8219  &   -28.9123 \\
\hline

\tablebreak

090523  &  &     \\
Center  &  22.8383  &  -62.2047 \\
Corners &  23.0738  &   -62.1859 \\
        &  23.2792  &   -62.2169 \\
        &  22.6023  &   -62.2230 \\
        &  22.3971  &   -62.1907 \\
\hline
100223  &  &    \\
Center  &  101.6711 &   +12.8302 \\
Corners &  101.9024 &    +12.6261 \\
        &  101.4091 &    +13.4941 \\
        &  101.5176 &    +12.9034 \\
        &  102.0085 &    +12.0774 \\
\hline
100629  &  &    \\
Center  &  227.6190  &  +29.5479 \\
Corners &  227.6295  &   +29.2934 \\
        &  227.5705  &   +29.7415 \\
        & 227.6081  &   +29.8012 \\
        & 227.6671  &   +29.3536 \\
\hline
100811  &  &    \\
Center  & 344.9605 & +20.6223 \\
Corners & 344.9395 &  +20.5411 \\
        & 345.0289 &  +20.6525 \\
        & 344.9816 &  +20.7032 \\
        & 344.8922 &  +20.5919 \\
\hline
101009  &  &    \\
Center  &   75.6068  &  -33.9325   \\
Corners &   75.8367  &  -33.8146   \\
        &   76.0209  &  -33.6624   \\
        &   75.1904  &  -34.1986   \\
        &   75.3765  &  -34.0492   \\
\hline
101129  &  &     \\
Center  &  268.6211  &   -7.7232   \\
Corners &  272.2014  &   -9.1377   \\
        &  265.1168  &   -6.9176   \\
        &  272.2546  &   -9.3668   \\
        &  265.1792  &   -7.0998   \\
\hline
110510  &  &     \\
Center  & 336.1407 & -44.0726 \\
Corners &  344.4202  &   -52.3621 \\
        & 344.5801   &  -52.3041 \\
        & 327.7177   &  -35.8320 \\
        & 327.8448  &   -35.7920 \\
\hline
\tablebreak
110705  &  &     \\
Center  &  156.0322   &  40.1169   \\
Corners &  156.1434   &  40.1805   \\
        &  156.0390   &  39.9001   \\
        &  156.0246   &  40.3325  \\
        &  155.9211   &  40.0532   \\
\hline
110802  &  &    \\
Center  &   44.4655  &   33.0037   \\
Corners &   44.4647  &   33.6264   \\
        &   44.4285  &   32.4871   \\
        &   44.5087  &   33.5009   \\
        &   44.4754  &   32.3503   \\
\hline
120222  &  &    \\
Center  & 296.9268  & 14.7102  \\
Corners &  290.7915  &   -5.6861 \\
        &  291.1995  &   -2.6858 \\
        &  304.9807  &   31.3395 \\
        &  307.1167  &   33.7741 \\
\hline
120519  &  &     \\
Center  &  178.0457  &   21.9477   \\
Corners &  178.2141  &   22.2660   \\
        &  178.2885  &   22.1813   \\
        &  177.8043  &   21.7128   \\
        &  177.8781  &   21.6273   \\
\hline
120811  &  &     \\
Center  &   43.7599  &  -31.9123   \\
Corners &   43.8831  &  -31.9728   \\
        &   43.7269  &  -32.0579   \\
        &   43.7928  &  -31.7664   \\
        &   43.6370  &  -31.8516   \\
\hline
121023  &  &     \\
Center  & 313.3540 & -4.6025 \\
Corners & 315.3258 &     -8.9120 \\
        & 311.5238 &     -0.2314 \\
        & 315.1761 &     -8.9571 \\
        & 311.3903 &     -0.3093 \\
\hline
121127  &  &     \\
Center  &  176.4314  &  -52.4316   \\
Corners &  176.2965  &  -52.4618   \\
        &  176.4549  &  -52.2522   \\
        &  176.4080  &  -52.6106   \\
        &  176.5661  &  -52.4011   \\
\hline
\tablebreak
130501  &  &     \\
Center  &  350.5102  &    16.7266  \\
Corners &  350.4295  &   16.8223   \\
        &  350.4890  &   15.7404   \\
        &  350.5512  &   17.6829   \\
        &  350.5910  &   16.6309   \\
\hline
130504B &  &     \\
Center  &  347.1594 &    -3.8325  \\
Corners &  347.0950 &    -3.8347  \\
        &  349.6913 &    -9.6653  \\
        &  345.9151 &    -0.2257  \\
        &  347.2238 &    -3.8303  \\
\hline
131126A &  &     \\
Center  &  202.8995 &    51.5581  \\
Corners &  202.9690 &    51.6343  \\
        &  203.1321 &    51.5592  \\
        &  202.6668 &    51.5564  \\
        &  202.8303 &    51.4819  \\
\hline
140414B &  & \\
Center  &  121.8042 &   -38.1211  \\
Corners &  121.2343 &   -35.5319  \\
        &  121.0519 &   -33.7849  \\
        &  122.7335 &   -42.2323  \\
        &  122.4453 &   -40.6251  \\
\hline
140807  &                           &                             \\
Center  &  188.8547 &  36.2259   \\
Corners &  188.9062 &   36.3820  \\
        &  188.7683 &  36.3184   \\
        &  188.9406 &  36.1333   \\
        &  188.8030 &  36.0695   \\
\hline
140906C &                            &                              \\
Center  &  314.9144  &   2.0606   \\
Corners &  314.8105  &   2.5705   \\
        &  314.9946  &   2.0906   \\
        &  314.8341  &   2.0305   \\
        &  315.0226  &   1.5373   \\
\hline
141011A &                             &                             \\
Center  &  257.9473  &   -9.6520   \\
Corners &  258.0072  &   -9.3553   \\
        &  257.7749  &   -9.3571   \\
        &  258.1189  &   -9.9487   \\
        &  257.8862  &   -9.9489   \\

\enddata
\end{deluxetable}


\clearpage

\begin{thebibliography}{}
\bibitem[Abdo et al.(2015)]{abdo2015} Abdo, A.~A., Abeysekara, A.~U., Alfaro, R., et al.\ 2015, Astroparticle Physics, 64, 4
\bibitem[Alexandreas et al. (1993)]{alex1993}D. E. Alexandreas, G. E. Allen, D. Berley, et al. 1993, Physical Review Letters 71, 2524
\bibitem[Amenomori et al. (1995)]{amenomori1995}M. Amenomori, Z. Cao, B. Z. Dai, et al. 1995, Proc. International Cosmic Ray Conference 2, 112
\bibitem[Barnacka et al. (2012)]{ba1}Barnacka, A., Glicenstein, J.-F., and Moderski, R. 2012 \prd~86, 043001
\bibitem[Bhat et al. (1980)]{bh1}Bhat, P., Gopalakrishnan, N., Gupta, S., Ramana Murthy, P., Sreekantan, B., and Tonwar, S. 1980 \nat~284, 433
\bibitem[Cline and Hong (1996)]{c1}Cline, D. \& Hong, W. 1996, Astroparticle Physics 5, 175
\bibitem[Czerny et al. (1996)]{c2}Czerny, B., et al. 2011, New Astronomy 16, 33
\bibitem[Cline et al. (1997)]{c3} Cline, D., et al. 1997, \apj~486, 169
\bibitem[Cline et al. (1999)]{c4} Cline, D., et al. 1999, \apj~527, 827
\bibitem[Cline et al. (2003)]{c5} Cline, D., et al. 2003, Astroparticle Physics 18, 531
\bibitem[Cline et al. (2005)]{c6} Cline, D., et al. 2005, \apj~633, L73
\bibitem[Czerny et al. (2011)]{c7} Czerny, B., et al. 2011, New Astronomy 16, 33
\bibitem[Cline and Hong (1992)]{c8} Cline, D. \& Hong, W. 1992, \apjl~401, L57
\bibitem[Eichler et al. (1989)]{e1} Eichler, D., Livio, M., Piran, T., and Schramm, D. 1989, \nat~340, 126
\bibitem[Fegan et al. (1978)]{fe1}Fegan, D., McBreen, B., O'Brien, D., and O'Sullivan, C. 1978, \nat~271, 731
\bibitem[Fichtel et al. (1994)]{fi1}Fichtel, C., Bertsch, D., Dingus, B., et al. 1994, \apj~434, 557
\bibitem[Frontera et al. (2009)]{f1}Frontera, F., Guidorzi, C., Montanari, E., et al. 2009, \apjs~180, 192												
\bibitem[Glicenstein et al. (2013)]{glicenstein2013} J. Glicenstein, A. Barnacka, M. Vivier, et al. 2013, arXiv:1307.4898
\bibitem[Goldstein et al. (2012)]{g2}Goldstein, A., Burgess, J. M., Preece, R., et al. 2012 \apjs~199, 19												
\bibitem[Goldstein et al. (2013)]{g1}Goldstein, A., Preece, R., Mallozzi, R., et al. 2013 \apjs~208, 21
\bibitem[Golenetskii et al. (2011a)]{g5}Golenetskii, S., Aptekar, R., Mazets, E., et al. 2011a, GCN Circ. 12110
\bibitem[Golenetskii et al. (2011b)]{g6}Golenetskii, , S., Aptekar, R., Frederiks, D., et al. 2011b, GCN Circ. 12111
\bibitem[Golenetskii et al. (2011c)]{g7}Golenetskii, S., Aptekar, R., Frederiks, D., et al. 2011c, GCN Circ. 12249
\bibitem[Golenetskii et al. (2011d)]{g8}Golenetskii, S., Aptekar, R., Frederiks, D., et al. 2011d, GCN Circ. 12271
\bibitem[Golenetskii et al. (2012a)]{g9}Golenetskii, S., Aptekar, R., Mazets, E., et al., 2012a GCN Circ. 13313
\bibitem[Golenetskii et al. (2012b)]{g10}Golenetskii, S., Aptekar, R., Frederiks, D., et al. 2012b, GCN Circ. 13315
\bibitem[Golenetskii et al. (2012c)]{g11}Golenetskii, S., Aptekar, R., Mazets, E., et al. 2012c, GCN Circ. 13620
\bibitem[Golenetskii et al. (2012d)]{g12}Golenetskii, S., Aptekar, R., Frederiks, D., et al. 2012d, GCN Circ. 13621
\bibitem[Golenetskii et al. (2012e)]{g13}Golenetskii, S., Aptekar, R., Mazets, E., et al. 2012e, GCN Circ. 13627
\bibitem[Golenetskii et al. (2012f)]{g14}Golenetskii, S., Aptekar, R., Mazets, E., et al. 2012f, GCN Circ. 14021
\bibitem[Golenetskii et al. (2012g)]{g15}Golenetskii, S., Aptekar, R., Frederiks, D. et al. 2012g, GCN Circ. 14022
\bibitem[Golenetskii et al. (2013a)]{g16}Golenetskii, S., Aptekar, R., Mazets, E., et al. 2013a, GCN Circ. 14561
\bibitem[Golenetskii et al. (2013b)]{g17}Golenetskii, S., Aptekar, R., Frederiks, D., et al. 2013b, GCN Circ. 14565
\bibitem[Golenetskii et al. (2013c)]{g18}Golenetskii, S., Aptekar, R., Pal'shin, V., et al. 2013c, GCN Circ. 15550
\bibitem[Golenetskii et al. (2013d)]{g19}Golenetskii, S., Aptekar, R., Frederiks, D., et al. 2013d, GCN Circ. 15551
\bibitem[Golenetskii et al. (2014a)]{g20}Golenetskii, S., Aptekar, R., Mazets, E., et al. 2014a, GCN Circ. 16801
\bibitem[Golenetskii et al. (2014b)]{g21}Golenetskii, S., Aptekar, R., Frederiks, D., et al. 2014b, GCN Circ. 16807
\bibitem[Golenetskii et al. (2014c)]{g22}Golenetskii, S., Aptekar, R., Pal'shin, V., et al. 2014c, GCN Circ. 16906
\bibitem[Golenetskii et al. (2014d)]{g23}Golenetskii, S., Aptekar, R., Frederiks, D., et al. 2014d, GCN Circ. 16907
\bibitem[Guiriec et al. (2010)]{g4}Guiriec, S., Briggs, M., Connaughton, V., et al. 2010, \apj~725, 225
\bibitem[Hawking (1974)]{h1} Hawking, S. 1974, \nat~248, 30
\bibitem[Hurley et al. (2002)]{h3}Hurley, K., Berger, E., Castro-Tirado, A., et al. 2002, \apj~567,447
\bibitem[Hurley et al. (2010)]{h2} Hurley, K., Rowlinson, A., Bellm, E., et al. 2010, \mnras~403, 342
\bibitem[Hurley et al. (2011a)]{h4}Hurley, K., Goldsten, J., Mitrofanov, I., et al. 2011a, GCN Circ. 12247
\bibitem[Hurley et al. (2011b)]{h5}Hurley, K., Goldsten, J., Mitrofanov, I., et al. 2011b, GCN Circ. 12269
\bibitem[Ishida et al. (2012)]{i1}Ishida, Y., Tashiro, M., Terada, Y., et al. 2012, GCN Circ. 14044
\bibitem[Jelley, Baird, and O'Mongain (1977)]{j1} Jelley, J., Baird, G., and O'Mongain, E. 1977, \nat~267, 499
\bibitem[Keane et al. (2012)]{ke1}Keane, E., Stappers, B., Kramer, M., and Lyne, A. 2012, \mnras~425, L71
\bibitem[Linton et al. (2006)]{linton2006}E. T. Linton, R. W. Atkins, H. M. Badran, et al. 2006, JCAP 1, 13
\bibitem[MacGibbon and Webber (1990)]{mw}MacGibbon, J.~H. and Webber, B.~R. 1990 \prd~41, 3052
\bibitem[MacGibbon et al.(2008)]{mcp} MacGibbon, J.~H., Carr, B.~J., \& Page, D.~N. 2008, \prd~78, 064043
\bibitem[Masetti et al. (2000)]{m2}Masetti, N., Palazzi, E., Pian, E., et al. 2000, GCN Circ. 720												
\bibitem[Mazets et al. (2008)]{m1} Mazets, E., Aptekar, R., Cline, T., et al. 2008, \apj~680, 545
\bibitem[Paciesas et al. (1999)]{pa1}Paciesas, W., Meegan, C., Pendleton, G., et al. 1999 \apjs~ 122, 465												
\bibitem[Paciesas et al. (2012)]{pa3}Paciesas, W., Meegan, C., von Kienlin, A., et al. 2012 \apjs~199, 18												
\bibitem[Pagani and Evans (2014)]{pe1}Pagani, C., and Evans, P. A. 2014, GCN Circ. 17134
\bibitem[Page and Hawking (1976)]{p1} Page, D. \& Hawking, S. 1976, \apj~206, 1
\bibitem[Page (1977)]{p2} Page, D. 1976, \prd~13, 198
\bibitem[Page et al.(2008)]{Page2008} Page, D.~N., Carr, B.~J., \& MacGibbon, J.~H.\ 2008, \prd, 78, 064044 
\bibitem[Pal'shin et al. (2013)]{pa2}Pal'shin, V., Hurley, K., Svinkin, D., et al. 2013, \apjs~207, 38												
\bibitem[Park et al. (2000)]{p3}Park, H., Williams, G., Perez, D., et al. 2000,	GCN Circ. 873
\bibitem[Pelassa and Meegan (2013)]{pe1}Pelassa, V., and Meegan, C. 2013, GCN Circ. 15573
\bibitem[Phinney and Taylor (1979)]{pt1}Phinney, S., and Taylor, J. 1979 \nat~277, 117
\bibitem[Porter and Weekes (1977)]{pw1}Porter, N., and Weekes, T. 1977 \apj~212, 224
\bibitem[Porter and Weekes (1978)]{pw2}Porter, N., and Weekes, T. 1978 \mnras~183, 205
\bibitem[Porter and Weekes (1979)]{pw3}Porter, N., and Weekes, T. 1979 \nat~277, 199
\bibitem[Price et al. (2000)]{p4}Price, P., Axelrod, T., and Schmidt, B. 2000,	GCN Circ. 898
\bibitem[Rees (1977)]{r1} Rees, M. 1977, \nat~266, 333
\bibitem[Sakamoto et al. (2012)]{s3}Sakamoto, A., Tashiro, M., Terada, Y., et al. 2012, GCN Circ. 13350
\bibitem[Singer et al. (2013)]{si1}Singer, L., Kasliwal, M., and Cenko, S. 2013, GCN Circ. 15572
\bibitem[Singer (2014)]{si2}Singer, L., private communication 2014
\bibitem[Svinkin et al. (2015a)]{s1} Svinkin, D., Hurley, K., Aptekar, R., et al. 2015a, \mnras~447, 1028
\bibitem[Svinkin et al. (2015b)]{s2} Svinkin, D., et al. 2015b, in preparation												
\bibitem[Terekhov et al. (1994)]{t1} Terekhov, O., Denisenko, D., Lobachev, V., et al. 1994, Astron. Lett. 20(3), 265												
\bibitem[Tesic (2012)]{veritas2012}G. Tesic and VERITAS Collaboration 2012, Journal of Physics Conference Series 375 052024
\bibitem[Ukwatta \& Wo{\'z}niak(2016)]{ukwatta2015} Ukwatta, T.~N., \& Wo{\'z}niak, P.~R.\ 2016, \mnras, 455, 703 
\bibitem[Ukwatta et al.(2015)]{ukwatta2015pbh} Ukwatta, T.~N., Stump, D., MacGibbon, J.~H., et al.\ 2015, arXiv:1510.04372
\bibitem[von Kienlin et al. (2014)]{v1}von Kienlin, A., Meegan, C., Paciesas, W., et al. 2014 \apjs~211, 13												
\bibitem[von Kienlin (2013)]{v2}von Kienlin, A. 2013, GCN Circ. 14560
\bibitem[von Kienlin (2014)]{v3}von Kienlin, A. 2014, GCN Circ. 16905
\bibitem[Vreeswijk and Rol (2000)]{v2}Vreeswijk, P., and Rol, E. 2000,	GCN Circ. 908
\bibitem[Xiong and Meegan (2012)]{x1}Xiong, S. Meegan C. 2012, GCN Circ. 13644
\bibitem[Yasuda et al. (2011)]{y1}Yasuda, T., Terada, Y., Tashiro, M., et al. 2011, GCN Circ. 12114
\bibitem[Yasuda et al. (2013)]{y2}Yasuda, T., Tashiro, Y., Terada, Y., et al. 2013, GCN Circ. 14600
\end{thebibliography}
\end{document}